\documentclass[12pt, a4paper]{article}
\pdfoutput=1
\usepackage{pstricks}
\usepackage{cite}
\usepackage{hyperref}
\usepackage{ dsfont }
\usepackage{amsmath}
\usepackage{amsfonts}
\usepackage{amssymb}
\usepackage{latexsym}
\usepackage{graphicx}
\usepackage{color}
\usepackage{enumerate}
\usepackage{multirow}
\usepackage{colortbl}
\usepackage{url}
\usepackage{enumitem}
\usepackage{comment}

 \def\be   {\begin{equation}}   \def\ee   {\end{equation}}
 \def\ba   {\begin{array}}      \def\ea   {\end{array}}
 \def\bea  {\begin{eqnarray}}   \def\eea  {\end{eqnarray}}
 \def\bean {\begin{eqnarray*}}  \def\eean {\end{eqnarray*}}
 \def\ga   {\gamma}
  \def\Ga   {\Gamma}
  \def\th   {\theta}
    \def\Th   {\Theta}
  \def\la   {\lambda}
    \def\La   {\Lambda}
  
  \def\ka{\kappa}
  \def\sig   {\sigma}

 \def\nn{\nonumber}
 \def\lee { \left( }
\def\rii { \right) }
\def\lan   {\langle}
\def\ran   {\rangle}
\def\de {\delta}
\def\De {\Delta}

\def\schi{\sin\chi}

\def\cw {\cos\theta_w}
\def\sw {\sin\theta_w}
\def\cwt {\cos^2\theta_w}
\def\swt {\sin^2\theta_w}
\def\cz {\cos\zeta}
\def\sz {\sin\zeta}
\def\tc {\tan\chi}
\def\sc {\sin\chi}

\def\secc {\sec\chi}
\def\svxr{\lan  \sig v \ran_{\mbox{\tiny{X-ray}}}}
\def\to {\rightarrow}

\newcommand{\sigmav}{\ensuremath{\langle\sigma v\rangle}}

\newcommand{\gev}{\ensuremath{\,\mathrm{GeV}}}

\newcommand{\kpc}{\ensuremath{\,\mathrm{kpc}}}
\newcommand{\kev}{\ensuremath{\,\mathrm{keV}}}

\numberwithin{equation}{section}

\begin{document}

\vspace{0cm}
\hfill IPMU14-0337 \\
\vspace{0cm}
\hfill LCTS/2014-45\\

\begin{center}
\vspace{2cm}

{\Large
Non-abelian Dark Matter Solutions for Galactic \\
Gamma-ray Excess and
Perseus 3.5 keV X-ray Line
}
\\ [2.5cm]
{\normalsize{\textsc{
Kingman Cheung$^{\,a,b,}$\footnote{\textsl{cheung@phys.nthu.edu.tw}}, Wei-Chih Huang$^{\,c,}$\footnote{\textsl{wei-chih.huang@ucl.ac.uk}}, Yue-Lin Sming Tsai$^{\,d,}$\footnote{\textsl{yue-lin.tsai@ipmu.jp}}}}}
\\[1cm]

\normalsize{\textit{
$^{a}$~Department of Physics, National Tsing Hua University, Hsinchu 300, Taiwan\\ \vspace{1.5mm}
$^{b}$~Division of Quantum Phases and Devices, School of Physics, 
Konkuk University, Seoul 143-701, Republic of Korea \\ \vspace{1.5mm}
$^{c}$~Department of Physics and Astronomy, University College London, UK\\ \vspace{1.5mm}
$^{d}$~Kavli IPMU (WPI), The University of Tokyo, Kashiwa, Chiba 277-8583, Japan
}}
\\ [1 cm]
{ \large{\textrm{
Abstract
}}}
\\ [0.5cm]
\end{center}

We attempt to explain simultaneously the Galactic center
gamma-ray excess and the 3.5 keV X-ray line from the Perseus cluster
based on a class of non-abelian $SU(2)$ DM models, in which
the dark matter and an excited state comprise a ``dark'' $SU(2)$ doublet.  The
non-abelian group kinetically mixes with the standard model gauge
group via  dimensions-5 operators.  
The dark matter particles annihilate into standard model fermions,
followed by fragmentation and bremsstrahlung, and thus
producing a continuous spectrum of gamma-rays.
On the other hand, 
the dark matter particles can annihilate into a pair of 
excited states, each of which decays
back into the dark matter particle and an X-ray photon, 
which has an energy equal 
to the mass difference between the dark matter and the excited state,
which is set to be 3.5 keV.
The large hierarchy
between the required X-ray and $\ga$-ray annihilation cross-sections 
can be achieved by a very small kinetic mixing between the SM and dark 
sector, 
which effectively suppresses the
annihilation into the standard model fermions but not into the excited
state.

\def\thefootnote{\arabic{footnote}}
\setcounter{footnote}{0}
\pagestyle{empty}

\newpage
\pagestyle{plain}
\setcounter{page}{1}

\section{Introduction}\label{sec:intro}

A gamma-ray excess around a few GeV near the Galactic center~(GC)
region, seen by the Fermi-LAT collaboration (see, for instance, the recent
analysis by the collaboration~\cite{Fermi-LAT:2014sfa}), has been
widely discussed based on dark matter~(DM) annihilations into standard
model~(SM) fermions ~\cite{Goodenough:2009gk,Hooper:2010mq,
Boyarsky:2010dr,Abazajian:2012pn, Gordon:2013vta,Huang:2013apa,Okada:2013bna,Modak:2013jya,
Abazajian:2014fta,Daylan:2014rsa,Lacroix:2014eea,Okada:2014usa,Zhou:2014lva,Wang:2014elb}, which hadronize
into neutral pions followed by $\pi^0 \to \gamma\gamma$, or
electromagnetic bremsstrahlung. 
On the other hand, recent reports of the $3.5\kev$ X-ray line~\cite{Bulbul:2014sua,Boyarsky:2014jta} from the XMM-Newton data
have triggered many studies in the context of DM, for example,
 Refs.~\cite{Cline:2014kaa,
Abazajian:2014gza,Adulpravitchai:2014xna,Baek:2014poa,Baek:2014qwa,Babu:2014pxa,Bezrukov:2014nza,Boddy:2014qxa,Bomark:2014yja,Chakraborty:2014tma,Chen:2014vna,Chiang:2014xra,Choi:2014tva,Cicoli:2014bfa,Cline:2014eaa,Conlon:2014xsa,Conlon:2014wna,Dubrovich:2014xaa,Dudas:2014ixa,Allahverdi:2014dqa,Dutta:2014saa,Finkbeiner:2014sja,Frandsen:2014lfa,Frigerio:2014ifa,Farzan:2014foo,Faisel:2014gda,Geng:2014zqa,Haba:2014taa,Higaki:2014zua,Higaki:2014qua,Ishida:2014dlp,Ishida:2014fra,Jaeckel:2014qea,Kong:2014gea,Kolda:2014ppa,Ko:2014xda,Lee:2014koa,Liew:2014gia,Modak:2014vva, Nakayama:2014ova,Nakayama:2014cza,Nakayama:2014rra,Okada:2014oda,Okada:2014zea,Patra:2014pga,Queiroz:2014yna,Robinson:2014bma,Rodejohann:2014eka}. 
Roughly speaking, they can be classified
into two categories: (i) DM undergoes upscattering into an excited stated 
followed by the decay back into DM and an X-ray photon; and (ii) 
decaying DM matter, such as a 7 keV sterile neutrino decaying into 
an active neutrino and the X-ray photon.
The excited DM, however, has a advantage of explaining some null results on X-ray line
searches due to a low local DM velocity as shown in Ref~\cite{Cline:2014vsa}.

There is a very interesting connection between the $\ga$-ray excess
and X-ray line as follows.  The GC $\ga$-ray excess can be explained
by annihilating DM with a mass from 10 to 60
GeV~\cite{Goodenough:2009gk,Abazajian:2012pn,Gordon:2013vta,Huang:2013apa,Okada:2013bna,
Abazajian:2014fta,Daylan:2014rsa,Lacroix:2014eea}, depending on the
final state of the annihilation.
On the other hand, due to the fact that the current DM
velocity is around $10^{-3}\, c$ in the Perseus cluster, where the
X-ray line is observed, the DM with a mass of 10 to 60 GeV 
coincidently has a kinetic energy of a few
keV. It implies if there exists an excited state with a 3.5 keV mass
splitting from the DM particle, then the DM particles can annihilate
into the excited state, followed by the decay back into the DM particle with a
photon accounting for the observed X-ray line.

In this work, we employ a class of non-abelian $SU(2)_X$ DM models
proposed in Refs.~\cite{Chen:2009ab,Cline:2014kaa}, where the DM
particle and the excited state form an $SU(2)_X$ doublet with a 3.5
keV mass splitting.  The $SU(2)_X$ kinetically mixes with the SM gauge
group via dimension-5 operators, through which the SM particles can couple
to the $SU(2)_X$ currents and the DM (and the excited state) couples to
the SM currents.  As mentioned above, the GC $\ga$-ray excess comes from
the DM annihilation into SM fermions accompanied by photon emission
while the X-ray line is realized
from the DM annihilation into the excited
state followed by the subsequent decay.  
Besides, the annihilation into SM
fermions, induced by the kinetic mixing, is suppressed compared to
that into the excited state if the kinetic mixing is small. This
suppression naturally explains the hierarchy between the required
annihilation cross-sections for the $\ga$-ray excess~($10^{-26}$
cm$^3$sec$^{-1}$) and the X-ray line emission 
($10^{-19}$ cm$^3$sec$^{-1}$) as shown
below. Note that similar ideas connecting the $\ga$-ray and X-ray
excess have been suggested in
Refs.~\cite{Finkbeiner:2014sja,Cline:2014kaa} with intermediate
states~(instead of the SM fermion final state) while an effort
connecting the $511$ keV line~\cite{Knodlseder:2005yq} and the GC
$\ga$-ray excess turns out to be negative~\cite{Boehm:2014bwa}.
  
This paper is organized as follows. In Sec.~\ref{sec:model}, we
specify the model and divide into the Majorana and Dirac DM cases.  In
Sec.~\ref{sec:X-sec}, we calculate the relevant cross-sections. In
Sec.~\ref{sec:observables}, we discuss the calculations of 
$\ga$-ray and X-ray flux as well as the DM relic abundance. 
In Sec.~\ref{sec:data analysis},
we present our numerical analysis with separation into the Majorana and 
Dirac cases.
Finally, we conclude in Sec.~\ref{sec:conclusion}.

\section{Non-abelian Dark Matter Models}\label{sec:model}

For the nonabelian DM model, we employ a ``dark'' $SU(2)_X$ gauge
group with kinetic mixing with the SM gauge groups proposed in
Refs.~\cite{Chen:2009ab,Cline:2014kaa}.  
We start with a $SU(2)_X$ doublet, which is comprised of the fields for 
the DM particle and an excited state.
In the following we will discuss two cases: $(i)$ Majorana DM~($\chi_1$)
with the Dirac excited state~($\psi_2$) and $(ii)$ Dirac DM~($\psi_1$)
with the Dirac excited state~($\psi_2$).\footnote{In this work, we
  denote Majorana particles by $\chi$ and Dirac particles with $\psi$
  for particles in the dark sector.} As we shall see later, we have
to make use of the resonance enhancement in order to achieve large
annihilation cross-sections, especially for explaining 
the X-ray line. The resonance
enhancement does not occur if both DM and the excited state are
Majorana with nearly degenerate masses, as shown in Appendix~\ref{reso_can}.
On the other hand, the Dirac DM with the Majorana excited state will lead
to a large $\ga$-ray flux but a small X-ray one, in contradiction to the $\ga$-ray and X-ray data
Therefore, we will not  discuss these two scenarios in this work.
The Lagrangian of the model reads,
\be
\mathcal{L} = \mathcal{L}_{SM} +  \mathcal{L}_{DM_1} +  \mathcal{L}_{DM_2}  
+ \mathcal{L}_{mix},
\ee
where $\mathcal{L}_{SM}$ is the SM Lagrangian.
$\mathcal{L}_{DM_{1,2}}$ correspond to the DM sector, including the DM 
doublet and the dark $SU(2)$ gauge bosons,
$X^{a}$~($a=(1,2,3)$), and dark Higgs triplets/doublets, which are used 
to provide masses to $\chi$s and $X$s:  
\be
\mathcal{L}_{DM_1} = -\frac{1}{4} X^{\mu\nu a}X^a_{\mu\nu} + \lee D^X_\mu \De_1 \rii^{\dag} (D^X_\mu \De_1)   + 
\lee D^X_\mu \De_2 \rii^{\dag} (D^X_\mu \De_2) ,
\ee
where $D^X_\mu$ is the covariant derivative of $SU(2)_X$ and
$\De_{1,2}$ are $SU(2)$ triplets, whose vacuum expectation values~(VEVs) provide masses to dark gauge 
bosons.
Note that one can play with the structure of $\lan \De_i \ran$ to give different masses to $X^a$.
For example, with $\lan \De_2 \ran = \lee 0 , v,  0 \rii^T$ in the isospin basis~(the first component has the highest isospin $I^X_3=1$,
the second with $I^X_3=0$, and so on) $X^{1,2}$ are massive but $X^3$ remains massless.

\subsection{Majorana DM}\label{sec:Majo_DM}
In the case of Majorana DM, the $\mathcal{L}_{DM_2}$ takes
the form
\be
\mathcal{L}_{DM_2} = i \chi^\dag D^X_\mu \sig^\mu \chi  +  i \tilde{\chi}_2^\dag \partial_\mu \bar{\sig}^\mu \tilde{\chi}_2 
+ \lee \frac{1}{2} \la_{\De} \lee  \chi \cdot \De_1 \cdot \chi \rii + \la_{h_2} \lee \chi \cdot h_D \rii \tilde{\chi}_2  + h.c. \rii,
\ee
where the two-component Weyl spinor notation is employed. 
Here ``$\cdot$'' refers to the $SU(2)$-invariant multiplication.
$\chi$ is an $SU(2)_X$ doublet,
consisting of two Weyl spinors, $\chi_1$ and $\chi_2$:  
$\chi=(\chi_2\,\,\, \chi_1)^T$. In addition, $h_D$ is
an $SU(2)_X$ scalar doublet. 
$\tilde{\chi}_2$ is a singlet under $SU(2)_X$, which will be paired up
with $\chi_2$ to form a Dirac fermion. The conversion between 
Dirac- and Weyl-spinors for $\chi_1$, $\chi_2$ and $ \tilde{\chi}_2$ is:
\bea
\psi_1&=& \left(
  \begin{array}{c}
    \chi_1 \\
    \chi^\dag_1 \\
  \end{array}
\right) , \nn\\
\psi_2&=&\left(
  \begin{array}{c}
    \chi_2 \\
    \tilde{\chi}^\dag_2 \\
  \end{array}
\right).
\eea
The corresponding $X^3$-current in the Weyl and Dirac-spinor notation is 
given by
\bea
\mathcal{L}  &\supset& g_X X^3_{\mu} J^{\mu}_X =
- \frac{g_X}{2} X^3_\mu \, \chi^\dag_1 \bar{\sigma}^\mu  \chi_1 +  \frac{g_X}{2} X^3_\mu \, \chi^\dag_2 \bar{\sigma}^\mu  \chi_2 \nn\\
&=& - \frac{g_X}{2} X^3_\mu \, \bar{\psi}_1 \ga^\mu \lee-\frac{\ga^5}{2}\rii  \psi_1 +
\frac{g_X}{2} X^3_\mu \,\bar{\psi}_2 \ga^\mu \lee\frac{1-\ga^5}{2}\rii  \psi_2,
\label{eq:Jx_Maj}
\eea
where the pre-factors $\pm 1/2$ come from the fact that 
$\chi_{2(1)}$ has $SU(2)_X$ isospin $1/2~(-1/2)$.

In order to give a Majorana mass to $\chi_1$, one can
make use of the lowest isospin~($I^X_3=-1$) component of $\lan \De_1 \ran$, leaving VEVs of other components vanishing,
i.e., $\lan \De_1 \ran = \lee 0 , 0,  v_{-1} \rii^T$ in the isospin basis.
The $\chi_1$ mass becomes $\la_\De  v_{-1}$.
Similarly, with the lower isospin~($I_3=-1/2$) of $\lan h_D \ran$, the Dirac mass of $\chi_2$ and $\tilde{\chi}_2$
becomes $\la_h v_{-1/2}$, where $v_{-1/2}$ is the VEV of the component of $I^X_3= -1/2$. Moreover,
$X^a$'s masses, at phenomenological level, are considered independent since as mentioned above one can always use $\lan \De_2 \ran$
to give a mass to specific gauge boson(s).

The particle content in the dark sector and the relevant quantum numbers 
in this model are summarized in Table~\ref{tab:quantum_number}.

\begin{table}[!h!]
\centering
\begin{tabular}{c c c c c c}
  \hline\hline
  Field & $\De_{1,2}$ & $h_D$ & $\chi$ & $\tilde{\chi}_2$ & $X^{1,2,3}$  \\
  \hline
 $SU(2)_X $ &  3 & 2 & 2 & 1 & 3 \\
 \hline
 spin &  0 & 0 &1/2 & 1/2 & 1 \\
 \hline\hline

\end{tabular}
\caption{\emph{The particle content and quantum numbers in the dark sector for the Majorana case. }}
\label{tab:quantum_number}
\end{table}
We would like to point out that the VEVs of $\De_{1,2}$ and $h_D$ are used to give a mass 
to the particles of interest and induce the kinetic mixing 
between the SM and the dark sector.
We simply assume that they are very heavy and play no roles in the 
context of GC gamma ray excess
and the 3.5 keV X-ray line.

\subsection{Dirac DM}\label{sec:Majo_DM}
In the case of Dirac DM, the $\mathcal{L}_{DM_2}$ takes
the form
\be
\mathcal{L}_{DM_2} = i \chi^\dag D^X_\mu \sig^\mu \chi  + \sum^2_{i=1} i \tilde{\chi}_i^\dag \partial_\mu \bar{\sig}^\mu \tilde{\chi}_i 
+ \lee \la_{h_1} \lee \chi \cdot h_{D_1} \rii \tilde{\chi}_1 + \la_{h_2} \lee \chi \cdot h_{D_2} \rii \tilde{\chi}_2  + h.c. \rii,
\ee
where $\lan h_{D_1} \ran=(v_1,0)^T$~($\lan h_{D_2}\ran =(0,v_2)^T$) gives a 
Dirac mass to $\chi_1$ and
$\tilde{\chi}_1$~($\chi_2$ and $\tilde{\chi}_2$). We list the particle content 
and quantum numbers in Table~\ref{tab:quantum_number_D}.
The conversion between Dirac- and Weyl-spinors for $\chi_1$, $\chi_2$ and 
$ \tilde{\chi}_2$ is:
\bea
\psi_1&=& \left(
  \begin{array}{c}
    \chi_1 \\
    \tilde{\chi}^\dag_1 \\
  \end{array}
\right) , \nn\\
\psi_2&=&\left(
  \begin{array}{c}
    \chi_2 \\
    \tilde{\chi}^\dag_2 \\
  \end{array}
\right),
\eea
and the corresponding $X^3$-current in the Weyl and Dirac-spinor notation is 
\bea
\mathcal{L}  &\supset& g_X X^3_{\mu} J^{\mu}_X =
- \frac{g_X}{2} X^3_\mu \, \chi^\dag_1 \bar{\sigma}^\mu  \chi_1 +  \frac{g_X}{2} X^3_\mu \, \chi^\dag_2 \bar{\sigma}^\mu  \chi_2 \nn\\
&=& - \frac{g_X}{2} X^3_\mu \, \bar{\psi}_1 \ga^\mu \lee\frac{1-\ga^5}{2}\rii  \psi_1 +
\frac{g_X}{2} X^3_\mu \,\bar{\psi}_2 \ga^\mu \lee\frac{1-\ga^5}{2}\rii  \psi_2.
\label{eq:Jx_Dir}
\eea

\begin{table}[!h!]
\centering
\begin{tabular}{c c c c c c}
  \hline\hline
  Field & $\De_{1,2}$ & $h_{D_{1,2}}$ & $\chi$ & $\tilde{\chi}_{1,2}$ & $X^{1,2,3}$  \\
  \hline
 $SU(2)_X $ &  3 & 2 & 2 & 1 & 3 \\
 \hline
 spin &  0 & 0 &1/2 & 1/2 & 1 \\
 \hline\hline

\end{tabular}
\caption{\emph{The particle content and quantum numbers for the Dirac case. }}
\label{tab:quantum_number_D}
\end{table}

\subsection{Kinetic Mixing}\label{sec:Kin_mix}

Finally, $\mathcal{L}_{mix}$ describes the 
mixing between the $SU(2)_X$ and SM gauge 
groups~\cite{Chen:2009ab,Cline:2014kaa} via dimension-5~(dim-5) operators:
\be
\mathcal{L}_{mix}= \sum^2_{i=1} \frac{1}{\La_i} \De_i^a X^{\mu\nu}_a Y_{\mu\nu} 
\sim  \sum^2_{i=1}  \frac{ \lan \De^a_i \ran}{\La_i} X^{\mu\nu}_a Y_{\mu\nu},
\ee 
where the corresponding $X^a$ mixes with the SM $\ga$ and $Z$ 
once $\De^a_i$ obtains a VEV.
In this work, we choose $\mathcal{L}_{mix}$ to be
\be
\mathcal{L}_{mix}
= - \frac{\schi}{2} X^{\mu\nu}_{3} Y_{\mu\nu}  - \frac{\sin\chi^\prime}{2} X^{\mu\nu}_{1} Y_{\mu\nu}  ,
\label{eq:su2mix} 
\ee 
which implies $X^1$ and $X^3$ mixes with SM neutral gauge bosons at tree level.
The reason why we include $X^1$ in the mixing is to enable the excited state
$\psi_2$ to decay into the DM and a photon to explain the 3.5 keV X-ray line. 
Moreover, we assume $\sin \chi^\prime \ll \schi$ for simplicity and neglect
the effect of $\sin \chi^\prime$ in diagonalizing the gauge boson mass 
matrix.\footnote{It is a legitimate assumption as long as the lifetime of the
excited state $\psi_2$ is less than 1 sec, thus having no influence on Big-Bang nucleosynthesis. }   
The relevant Lagrangian, with Lorentz indices suppressed, 
before and after diagonalizing the mass matrix of $\ga$, $Z$ and $X^3$ reads  
\be
\mathcal{L} \supset
\left(
\begin{array}{ccc}
  A_f & Z_f  & X^3_f   
\end{array}
\right)
\left(
\begin{array}{c}
  e J_{EM} \\
  g J_Z  \\
 g_X J_X   
\end{array}
\right)
=
\left(
\begin{array}{ccc}
  A_m & Z_m  & X^3_m   
\end{array}
\right)
R
\left(
\begin{array}{c}
  e J_{EM} \\
  g J_Z  \\
 g_X J_X   
\end{array}
\right),
\ee
where $e$, $g$ and $g_X$ are $U(1)_{EM}$, $SU(2)_L$ and $SU(2)_X$ gauge 
couplings, respectively.
The subscript $f$ refers to the flavor states,
$m$ denotes the mass and kinetic eigenstates, and $J$s are currents.\footnote{To be more precise,
$J^\mu_{EM}= Q_f \bar{f} \ga^\mu f$, while
$J_Z=  \frac{1}{\cw}\bar{f} \ga^\mu  \lee \lee I_3 - \swt Q_f \rii P_L + \lee - \swt Q_f \rii P_R \rii f$
for a fermion $f$. $P_{L(R)}$ is the left-~(right-)handed projection operator.
$J_X$ are defined in Eq.~(\ref{eq:Jx_Maj}) and (\ref{eq:Jx_Dir})
for the Majorana and Dirac DM, respectively. } 
$R$ is the rotation matrix connecting the flavor and mass basis of 
the gauge bosons~\cite{Cassel:2009pu}:
\be
R=\left(
\begin{array}{ccc}
1  & 0  & 0  \\
-\cw\tc\sz  & \sw\tc\sz +\cz   & \secc\sz  \\
-\cw\tc\cz  & \sw\tc\cz -\sz  & \secc\cz  
\end{array}
\right),
\ee
where
\bea
\tan\lee 2 \zeta\rii&=& \frac{2 \de_X  \lee m^2_{X^3} - m^2_W \sec^2\th_w \rii}
{ \lee m^2_{X^3} - m^2_W \sec^2\th_w \rii^2 - \de^2_X  },  \nn\\
\de_X&=& - \frac{m^2_W \sw\tc}{\cwt}.
\label{eq:zeta_def}
\eea
It is clear that $R=\mathds{1}_{3\times 3} $ if $\sc=0$. Note that the 
photon does not
couple to $J_X$ at tree-level but the interaction will be induced at 
loop-level. From now on, we will suppress the subscript $m$ in the gauge 
bosons: $A$, $Z$ and $X_3$
refer to the mass and kinetic eigenstates, unless otherwise stated.

\section{Relevant annihilation cross-sections}\label{sec:X-sec}
\begin{figure}
   \centering
   \includegraphics[scale=0.85]{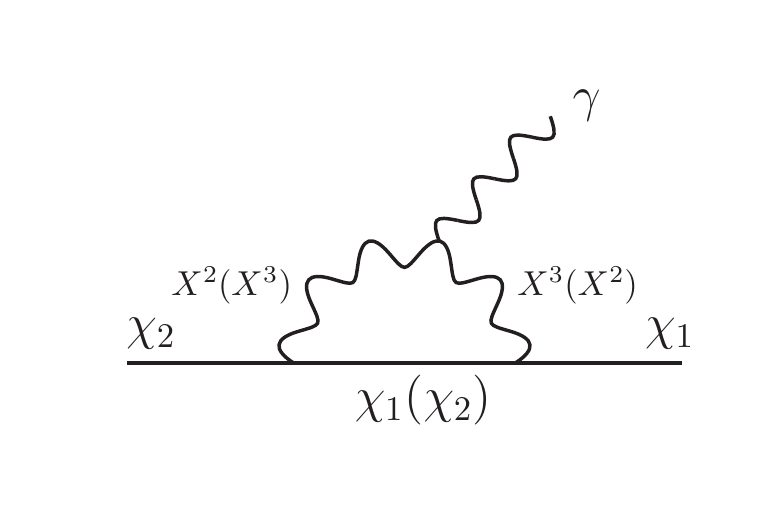}
 \caption{\emph{$\chi_2$ decays into $\chi_1$ and a photon through a 
dim-5 operator, $X^{\mu\nu}_1Y_{\mu\nu}$.}}
 \label{fig:x2_dec}
\end{figure}

In this section, we calculate the DM annihilation cross-sections 
into SM fermions and the excited state $\psi_2$.
The first process will give rise to $\ga$-rays via 
fragmentation of quarks and final state radiation from leptons,
while the second one will yield X-rays when
$\psi_2$ decays back into the DM and a photon 
via $\sin\chi^\prime X^{\mu\nu}_1 Y_{\mu\nu}$ as shown in Fig.~\ref{fig:x2_dec}.
 
In this work, we focus on the regime, where $m_{X^a} > m_{DM}$, 
such that the GC gamma-ray excess and 3.5 keV X-ray line can be 
realized through DM annihilations into SM particles and excited $\chi_2$,
respectively. As we shall see below, we need a large resonance 
enhancement in the annihilation cross-section coming from the $X^3$ narrow 
width; therefore, to a very good approximation, we only include 
$X^3$-exchange processes in the computation.

\subsection{Majorana DM}\label{sec:Majo_DM_Xsection}
For Majorana DM, we have the following relevant annihilation cross-sections:
$\chi_1 \chi_1 \to \bar{f} f$~($P$-wave) for $\ga$-ray and the DM density, $\chi_1 \chi_1 \to \bar{\psi}_2 \psi_2$~($P$-wave),
$ \bar{\psi}_2 \psi_2 \to \bar{f} f$~($S$-wave) for the DM density. In order to account for the X-ray line, 
the mass splitting between $m_{\chi_1}$ and $m_{\psi_2}$ is set to be 3.5 keV, which in turn implies that the $S$-wave
$\bar{\psi}_2 \psi_2 \to \bar{f} f$ is the dominant contribution to 
the DM abundance computation as opposed to
the $\ga$-ray excess and X-ray line, which arise from $P$-wave processes 
due to axial-vector interactions of $\chi_1$.
 
\begin{figure}
   \centering
   \includegraphics[scale=0.5]{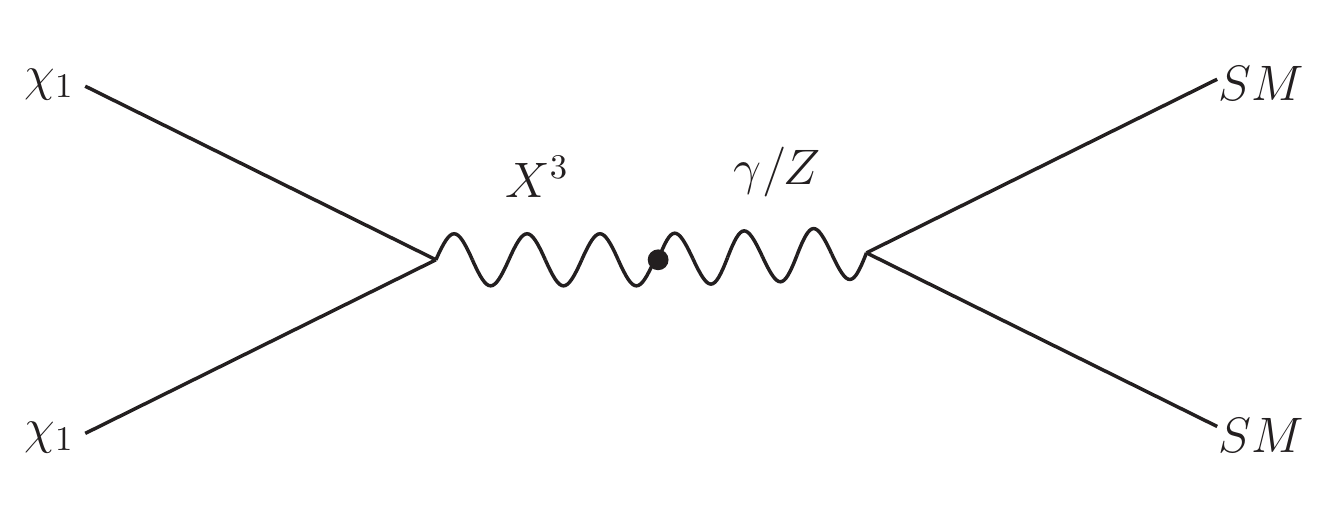}
 \caption{\emph{$\chi_1 \chi_1$ annihilate into SM particles 
that fragment into photons, which are responsible for GC gamma rays.}}
 \label{fig:GC-ga}
\end{figure}

For $\chi_1$ annihilating into SM fermions $f$ of mass $m_f$ via $X_3$, 
as shown in Fig.~\ref{fig:GC-ga}, the relevant interactions are
\footnote{Again, we use the Weyl spinor notation for Majorana $\chi_1$.}
\be
\mathcal{L}_{\chi_1\chi_1 \to \bar{f}f } \supset -\frac{1}{2}\lee g_X \secc\cz  \rii X^3_{\mu} \chi^\dag_1 \bar{\sig}^\mu \chi_1 + 
X^3_{\mu} \bar{f} \ga^{\mu} \lee g_L P_L + g_R P_R \rii f,
\label{eq:Xsec_xray}
\ee
where
\bea
g_L &=& - e Q_f \cw\tc\cz  +  \lee \sw\tc\cz -\sz \rii \frac{g}{\cw} \lee I_3 - \swt  Q_f \rii, \nn\\
g_R &=& - e Q_f \cw\tc\cz  +  \lee \sw\tc\cz -\sz \rii \frac{g}{\cw} \lee - \swt  Q_f \rii, \nn\\ 
\label{eq:gLgR}
\eea
in which $Q_f$ is the fermion electric charge and $I_3$ is the isospin, 
associated with left-handed field. 
The annihilation cross-section times the relative velocity $v$ is,
\be
\lee \sig v \rii_{\chi_1\chi_1\to \bar{f} f}
= \sum_f \frac{ \lee g_X \secc\cz  \rii^2   \sqrt{s -4 m^2_f} }{ 48 \pi \, m^4_{X^3} \, s^{3/2} \lee \lee s -m^2_{X^3} \rii^2 + \Ga^2_{X^3} m^2_{X^3} \rii} 
\lee \la_{\sig_1} + \la_{\sig_2} - \la_{\sig_3} \rii,
\label{eq:x1x1_ff}
\ee
where
\bea
s&=& \frac{ 4 m^2_{\chi_1} }{ 1 - v^2/4  }, \nn \\
\la_{\sig_1}&=&  s^2 \lee m^4_{X^3} \lee g^2_L  +g^2_R \rii + 6 m^2_{\chi_1} m^2_f \lee g_L -g_R   \rii^2 \rii , \nn \\
\la_{\sig_2}&=&   2 m^4_{X^3} m^2_{\chi_1} m^2_f   \lee 5 g^2_L - 18 g_L g_R + 5 g^2_R   \rii, \nn \\
\la_{\sig_3}&=&   s \, m^2_{X^3} \lee 4 m^2_{\chi_1}  \lee m^2_{X^3} \lee g^2_L + g^2_R \rii
+ 3 m^2_f \lee g_L -g_R   \rii^2 \rii + m^2_{X^3} m^2_f   \lee  g^2_L - 6 g_L g_R +  g^2_R  \rii \rii.  \nn
\eea
Note that one has to sum over all different final states as denoted by $\sum_f$.
 To simplify the expression, we employ the resonant limit of $m_{\chi_1}=\frac{1}{2} m_{X^3}$
on the matrix element while keeping the kinetic part intact.\footnote{We apply this simplification
to annihilation cross-sections below as well.}
The annihilation cross-section reads, up to the second order in $v^2$,
\be
\lee \sig v \rii_{\chi_1\chi_1\to \bar{f} f}
\simeq \sum_f \frac{ \lee g_X \secc\cz  \rii^2   \sqrt{s -4 m^2_f} }{ 96 \pi \,  \lee \lee s -m^2_{X^3} \rii^2 + \Ga^2_{X^3} m^2_{X^3} \rii }
m_{\chi_1}\lee g_1 v^2 + g_2 v^4   \rii ,
\label{eq:x1x1_ff_sim}
\ee
with
 \bea
g_1&=& \lee g^2_L + g^2_R \rii -  \frac{ m^2_f \lee g^2_L - 6 g_L g_R + g^2_R \rii }{4  m^2_{\chi_1} } , \nn \\
g_2&=& \frac{1}{8} \lee \lee  g^2_L + g^2_R \rii  +  \frac{ m^2_f \lee g^2_L - 3 g_L g_R + g^2_R \rii }{  m^2_{\chi_1}} \rii. \nn
\eea
Similarly, for $\bar{\psi}_2 \psi_2 \to \bar{f} f$, 
which is relevant for the relic abundance computation, we have,  up to the first order in $v^2$,
\be
\lee \sig v \rii_{ \bar{\psi}_2 \psi_2 \to \bar{f} f }
\simeq \sum_f \frac{ \lee g_X \secc\cz  \rii^2   \sqrt{s -4 m^2_f} }{ 64 \pi \,  \lee \lee s -m^2_{X^3} \rii^2 + \Ga^2_{X^3} m^2_{X^3} \rii }
m_{\psi_2}\lee h_1 + h_2 v^2   \rii ,
\label{eq:p2p2_ff_sim}
\ee
with
 \bea
 s&=& \frac{ 4 m^2_{\psi_2} }{ 1 - v^2/4  }, \nn \\
h_1&=& \lee g^2_L + g^2_R \rii -  \frac{ m^2_f \lee g^2_L - 6 g_L g_R + g^2_R \rii }{4  m^2_{\psi_2} } , \nn \\
h_2&=& \frac{5}{24}  \lee  g^2_L + g^2_R \rii  +   \frac{ m^2_f \lee g^2_L - 6 g_L g_R + g^2_R \rii }{  96 m^2_{\psi_2}}. \nn
\eea

In addition, the $X^3$ decay width $\Ga_{X^3}$ is given by, 
including the channels into $\chi_1$, $\psi_2$ and SM fermions,
\be
\Ga_{X^3} = \Ga_{X^3 \to \chi_1 \chi_1} +  \Ga_{X^3 \to \bar{\psi}_2 \psi_2}  +  \sum_f  \Ga_{X^3 \to \bar{f} f},
\ee
where
\bea
 \Ga_{X^3 \to \chi_1 \chi_1}&=&\Th \lee m_{X^3} - 2 m_{\chi_1} \rii \lee g_X \secc\cz  \rii^2  \frac{\lee m^2_{X^3} - 4 m^2_{\chi_1} \rii^{3/2} }{ 96 \pi \, m^2_{X^3} }, \nn\\
  \Ga_{X^3 \to \bar{\psi}_2 \psi_2}&=&\Th \lee m_{X^3} - 2 m_{\psi_2} \rii \lee g_X \secc\cz  \rii^2  \frac{\lee m^2_{X^3} -  m^2_{\psi_2} \rii 
  \lee m^2_{X^3} - 4 m^2_{\psi_2} \rii^{1/2}   }{ 96 \pi \, m^2_{X^3} }, \nn\\
   \Ga_{X^3 \to \bar{f} f}&=&\Th \lee m_{X^3} - 2 m_f \rii  \sqrt{ m^2_{X^3} -  4 m^2_f  } \,
    \frac{ m^2_{X^3} \lee g^2_L  + g^2_R \rii  -  m^2_f \lee g^2_L -6 g_L g_R + g^2_R  \rii   }{24 \pi \, m^2_{X^3} }. \nn\\
\eea

\begin{figure}
   \centering
   \includegraphics[scale=0.5]{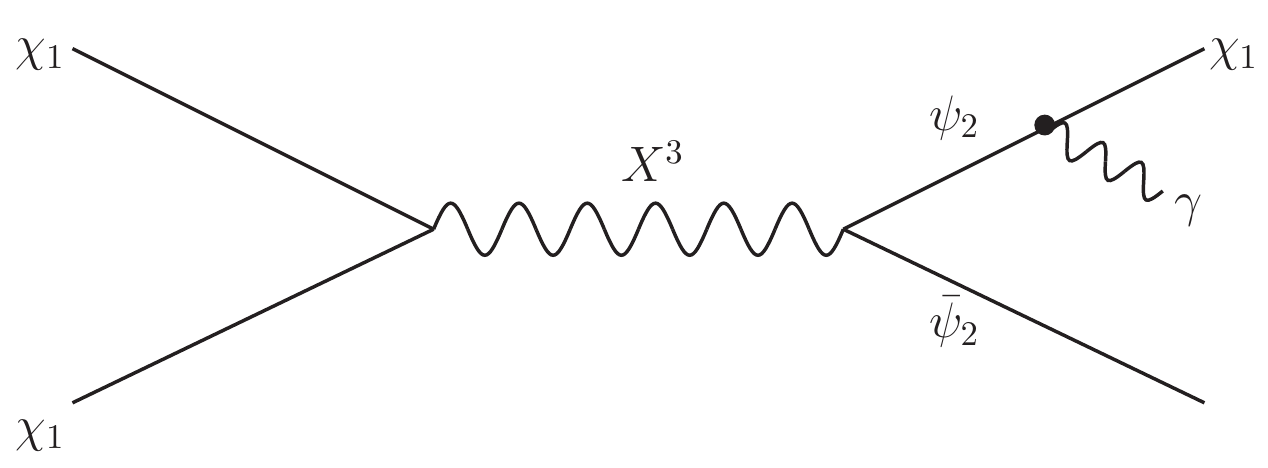}
 \caption{\emph{$\chi_1 \chi_1$ annihilation into $\bar{\psi}_2\psi_2$ 
responsible for the X-ray line.}}
 \label{fig:Xray}
\end{figure}

On the other hand, $\chi_1 \chi_1 \to \bar{\psi}_2 \psi_2$ 
with subsequent decay of $\psi_2$~(or $\bar{\psi}_2$) into $\chi_1$ and 
$\gamma$ explaining the 3.5 keV X-ray line, as shown in Fig.~\ref{fig:Xray},
has the cross-section
\be
\lee \sig v\rii_{\chi_1\chi_1 \to \bar{\psi}_2 \psi_2} = 
\frac{ \lee g_X \secc\cz  \rii^4   \sqrt{s -4 m^2_{\psi_2}} }{ 192 \pi \, m^4_{X^3} \, s^{3/2} \lee \lee s -m^2_{X^3} \rii^2 + \Ga^2_{X^3} m^2_{X^3} \rii} 
\lee \ka_{\sig_1} + \ka_{\sig_2} \rii,
\ee
where
\bea
\ka_{\sig_1}&=& 6 s^2  m^2_{\chi_1} m^2_{\psi_2}  -12\, s\, m^2_{X^3} m^2_{\chi_1} m^2_{\psi_2}  , \nn \\
\ka_{\sig_2}&=&   m^4_{X^3}\lee  2 m^2_{\chi_1}  \lee 5 m^2_{\psi_2} - 2\,s \rii  + s\lee s- m^2_{\psi_2} \rii  \rii. \nn \\
\eea
For $m_{\chi_1} \simeq m_{\psi_2}$, 
we have to a very good approximation, up to the second order in $v^2$:
\be
\lee \sig v\rii_{\chi_1\chi_1 \to \bar{\psi}_2 \psi_2}  \simeq 
\frac{ \lee g_X \secc\cz  \rii^4   \sqrt{s -4 m^2_{\psi_2}} }{  1536 \pi \,  \lee \lee s -m^2_{X^3} \rii^2 + \Ga^2_{X^3} m^2_{X^3} \rii }  m_{\chi_1} \lee v^2 \lee 3 + v^2 \rii \rii.
\label{eq:x1x1_x2x2_sim}
\ee
 
\subsection{Dirac DM}\label{sec:Dirac_DM_Xsection}
For Dirac DM, the distinctive feature compared to the Majorana case is that all relevant processes are $S$-wave dominated
due to the vector interactions of $\psi_1$. 
The annihilation cross-section of $ \bar{\psi}_1 \psi_1 \to \bar{f} f$ is,  up to the first order in $v^2$,
\be
\lee \sig v \rii_{ \bar{\psi}_1 \psi_1 \to \bar{f} f }
\simeq \sum_f \frac{ \lee g_X \secc\cz  \rii^2   \sqrt{s -4 m^2_f} }{ 64 \pi \,  \lee \lee s -m^2_{X^3} \rii^2 + \Ga^2_{X^3} m^2_{X^3} \rii }
m_{\psi_1}\lee \omega_1 + \omega_2 v^2   \rii ,
\label{eq:p1p1_ff}
\ee
with
 \bea
 s&=& \frac{ 4 m^2_{\psi_1} }{ 1 - v^2/4  }, \nn \\
\omega_1&=& \lee g^2_L + g^2_R \rii -  \frac{ m^2_f \lee g^2_L - 6 g_L g_R + g^2_R \rii }{4  m^2_{\psi_1} } , \nn \\
\omega_2&=& \frac{5}{24}  \lee  g^2_L + g^2_R \rii  +   \frac{ m^2_f \lee g^2_L - 6 g_L g_R + g^2_R \rii }{  96 m^2_{\psi_1}}. \nn
\eea
and for $ \bar{\psi}_1 \psi_1 \to \bar{\psi}_2 \psi_2$ we have
\be
\lee \sig v\rii_{\bar{\psi}_1\psi_1 \to \bar{\psi}_2 \psi_2}  \simeq 
\frac{ \lee g_X \secc\cz  \rii^4   \sqrt{s -4 m^2_{\psi_2}} }{  393216 \pi \,  \lee \lee s -m^2_{X^3} \rii^2 + \Ga^2_{X^3} m^2_{X^3} \rii }  m_{\psi_1} \lee 1152+ 336 v^2 + 83 v^4  \rii.
\ee
Note that the $X^3$ partial decay width into $\bar{\psi}_1 \psi_1$ becomes
\be
 \Ga_{X^3 \to \bar{\psi}_1 \psi_1}=\Th \lee m_{X^3} - 2 m_{\psi_1} \rii \lee g_X \secc\cz  \rii^2  \frac{\lee m^2_{X^3} -  m^2_{\psi_1} \rii 
  \lee m^2_{X^3} - 4 m^2_{\psi_1} \rii^{1/2}   }{ 96 \pi \, m^2_{X^3} }
\ee 

Furthermore, the Dirac DM $\psi_1$ will have sizable 
DM-nucleon  interactions in the context of direct detections. 
The effective DM-quark interaction reads,
\be
\mathcal{L} \supset  -\frac{\lee g_X \secc\cz  \rii \lee g_L  + g_R  \rii}{8 m^4_{X^3}}
\bar{\psi}_1 \ga^{\mu} \psi_1 \bar{f} \ga_{\mu}f,
\ee 
where $g_L$ and $g_R$ are defined in Eq.~\ref{eq:gLgR}.

\section{Observables}\label{sec:observables}
Based on the DM annihilation cross-sections into the excited state
and SM fermions, 
we now describe how to compute the flux of $X$-rays and $\ga$-rays,
and will comment on the DM relic density computation.  

\subsection{X-ray}\label{sec:x-ray}
Recently, a potential signal of a monochromatic photon line from the Perseus cluster 
at energy around $3.56\kev$ has been identified from 
the XMM-Newton data~\cite{Bulbul:2014sua,Boyarsky:2014jta}. 
The flux of such a monochromatic photon line at the X-ray energy 
$E_\gamma=3.56\kev$ is measured to be 
$\Phi_{\gamma\gamma}=5.2^{+3.70}_{-2.13}\times 10^{-5}$ ph cm$^{-2} s^{-1}$~\cite{Boyarsky:2014jta}. 
Although the source of this X-ray line signal is still unclear, the 
DM annihilation (or decay) into photons 
is a well motivated possibility~\cite{Cline:2014kaa,
Abazajian:2014gza,Adulpravitchai:2014xna,Baek:2014poa,Baek:2014qwa,Babu:2014pxa,Bezrukov:2014nza,Boddy:2014qxa,Bomark:2014yja,Chakraborty:2014tma,Chen:2014vna,Chiang:2014xra,Choi:2014tva,Cicoli:2014bfa,Cline:2014eaa,Conlon:2014xsa,Conlon:2014wna,Dubrovich:2014xaa,Dudas:2014ixa,Dutta:2014saa,Finkbeiner:2014sja,Frandsen:2014lfa,Frigerio:2014ifa,Farzan:2014foo,Faisel:2014gda,Geng:2014zqa,Haba:2014taa,Higaki:2014zua,Higaki:2014qua,Ishida:2014dlp,Ishida:2014fra,Jaeckel:2014qea,Kong:2014gea,Kolda:2014ppa,Ko:2014xda,Lee:2014koa,Liew:2014gia,Modak:2014vva, Nakayama:2014ova,Nakayama:2014cza,Nakayama:2014rra,Okada:2014oda,Okada:2014zea,Patra:2014pga,Queiroz:2014yna,Robinson:2014bma,Rodejohann:2014eka}. 
Considering the Perseus Mass $\simeq 1.49\times 10^{14} M_{\odot}$ and the 
distance between the Perseus cluster and the solar system $\simeq 78$ Mpc, 
the photon-line flux from DM annihilation can be written as  
\begin{equation}
    \frac{\Phi_{\gamma\gamma}}{\rm{cm}^{-2} s^{-1}} = 2.08\times 10^{-3}\times 
    \left[\frac{1\gev}{m_{\rm{DM}}}\right]^2\times 
    \frac{\mathcal{D}\sigmav_{\gamma\gamma}}{10^{-19}\rm{cm}^3s^{-1}},
    \label{eq:Xrayflux}
\end{equation} 
where $\mathcal{D}$ is 1 for Majorana DM and 1/2 for Dirac DM.
The monochromatic annihilation cross section 
$\sigmav_{\gamma\gamma}$ is the relative velocity averaged with 
all the DM inside the Perseus cluster. Here, we adopt the relative 
velocity $v_{rel.}$ described by 
the Maxwell-Boltzmann distribution~\cite{Ferrer:2013cla}, 
\begin{equation}
    f(v_{rel.})=\frac{v_{rel.}^2}{2\sqrt{\pi}v_0^3} 
    \exp\left[  \frac{-v_{rel.}^2}{4v_0^2} \right], 
    \label{eq:fv}
\end{equation} 
where we take the mean value of the velocity dispersion $v_0\sim 10^{-3}\, c$~\cite{Ferrer:2013cla}.    
One can see that a DM mass $m_{\rm{DM}}\sim 10\gev$ requires 
$\sigmav_{\gamma\gamma}\sim 2.5\times 10^{-19}\rm{cm}^3s^{-1}$ 
in order to explain the X-ray signal from the Perseus cluster.  

It is worthy to mention that the information of Perseus mass, which 
is constrained by the velocity dispersion, can substantially
reduce the uncertainties 
arising from halo inner slope. 
In Ref.~\cite{Finkbeiner:2014sja}, an overall uncertainty about a factor of 5 
was obtained for the DM flux predicted in Eq.~(\ref{eq:Xrayflux}).

\subsection{GC $\gamma$-ray}\label{sec:gamma-ray}

A gamma-ray excess in the GC region, 
found in the Fermi-LAT data, has been widely studied 
in the context of DM annihilation~\cite{Goodenough:2009gk,Hooper:2010mq,
Boyarsky:2010dr,Abazajian:2012pn, Gordon:2013vta,Huang:2013apa,
Abazajian:2014fta,Daylan:2014rsa,Lacroix:2014eea}. 
Assuming spherical symmetry, the spatial distribution of 
such an excess can be explained by DM annihilation in
the generalized Navarro-Frenk-White (gNFW,~\cite{Navarro:1996gj,Zhao:1995cp}) profile, 
\begin{equation}
    \rho(r) = \rho_s \frac{(r/r_s)^{-\gamma}}{(1 + r/r_s)^{3 - \gamma}}.
    \label{eq:NFWgen}
\end{equation}
To explain the gamma-ray excess, the inner slope $\gamma$ parameter 
requires $\gamma=1.2$~\cite{Daylan:2014rsa,Hooper:2013rwa}. 
In this work, we adopt this value together with the local density 
$\rho(8.5\kpc)=0.4\gev/\rm{cm}^3$ and $r_s=20\kpc$.

The differential diffuse gamma-ray flux along a line-of-sight (l.o.s.) 
at an open angle relative to the direction of the GC is given by
\begin{equation}
  \frac{dN}{dE}= \frac{\sigmav_\gamma}{\mathcal{D}\pi \,m_\chi^2} \, \frac{dN_{\gamma}}{dE}
  \int_\text{l.o.s.}\!\!\!\!\! ds\;\rho^2(r(s, \psi))\,,
  \label{eqn:fluxADM}
\end{equation}
where $\mathcal{D}$ is 8 for Majorana DM but 16 for Dirac one. 
The $\sigmav_\gamma$ is the velocity averaged annihilation cross section at the GC. 
However, the mean value of the velocity dispersion $v_0$ in Eq.~(\ref{eq:fv}) is $\sim 10^{-4}\,c$ 
at the GC region \cite{Ferrer:2013cla,Fornasa:2013iaa}. 
 
The $\frac{dN_{\gamma}}{dE}$ is the photon energy distribution per annihilation. 
All possible annihilation channels are included.
The branching ratio of all the possible annihilation channels can be obtained    
by using Eq.~\eqref{eq:x1x1_ff_sim} and \eqref{eq:p1p1_ff}. For each annihilation channel, 
the corresponding $\frac{dN_{\gamma}}{dE}$ is taken from the numerical \texttt{PPPC4} table~\cite{Cirelli:2010xx}.

One has to bear in mind that the background uncertainties for the GC gamma ray excess 
can significantly change the DM parameter space. 
Therefore, in order to include the background uncertainties, 
we use the central values and error bars in Fig. 17 from Ref.~\cite{Calore:2014xka}, 
where the systematic uncertainties coming from the Galactic diffuse emission 
have been properly included.  
Following Ref.~\cite{Calore:2014xka}, the inner Galactic central region described by  
the Galactic longitude $l$ and latitude $b$ is 
\begin{equation}
    |\ell|\leq 20^\circ \quad \text{and} \quad 2^\circ\leq |b| \leq 20^\circ\;.
    \label{eqn:ROI}
\end{equation}

 \begin{figure}
   \centering
   \includegraphics[scale=0.5]{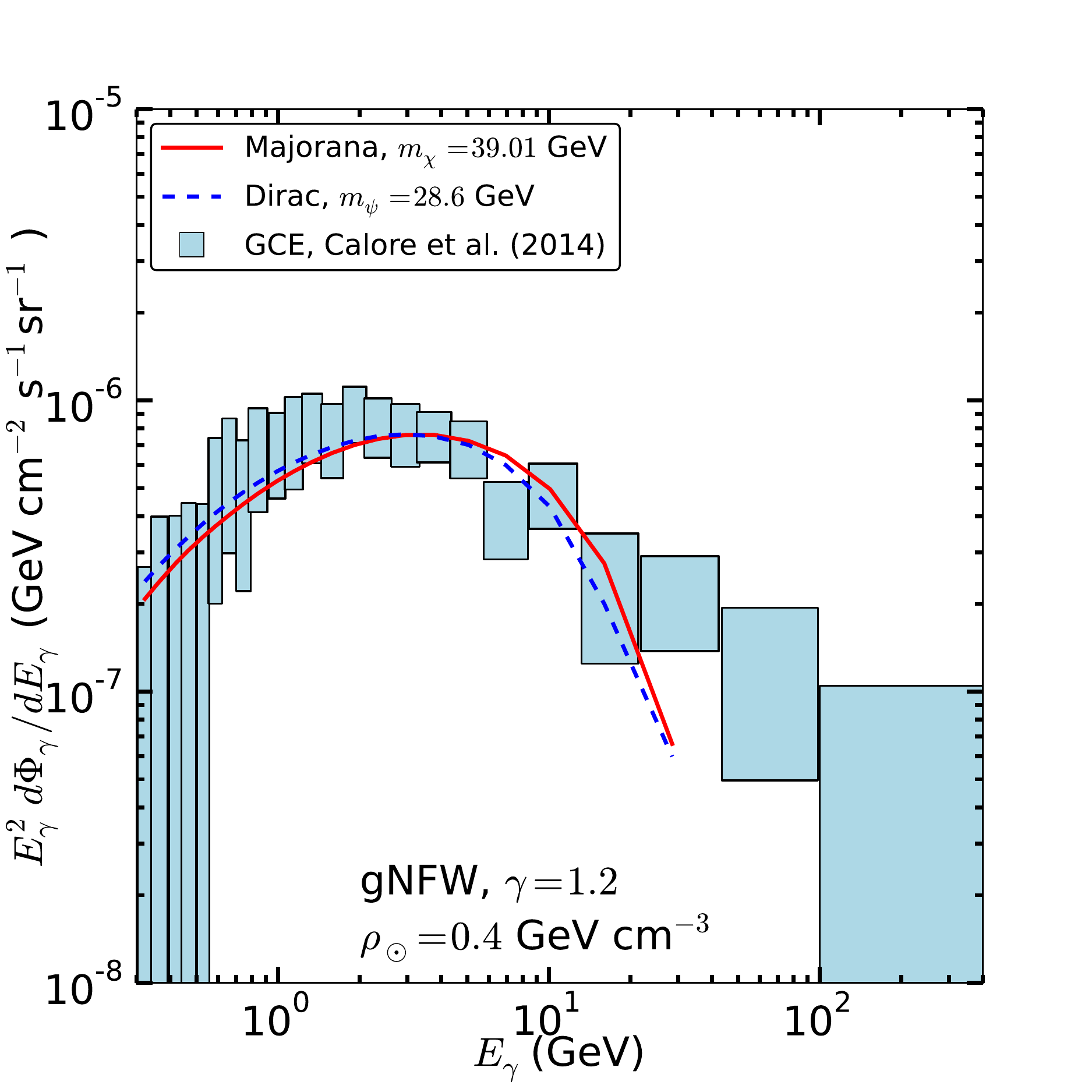}
 \caption{\emph{GC $\ga$-ray excess spectrum taken from Ref.~\cite{Calore:2014xka}. 
We also show the corresponding photon spectra obtained for the
 Majorana~(solid red line) and Dirac~(dashed blue line) case.}}
 \label{fig:GC_calore}
\end{figure}

We conclude this section with Fig.~\ref{fig:GC_calore} where the
data on $\ga$-ray spectrum is taken from Ref.~\cite{Calore:2014xka} and 
the photon spectra are calculated using our
best-fit points in both the Majorana~(solid red line) and Dirac cases~(dashed blue line), 
for which we include the $\ga$-ray and X-ray data into
fitting. One can see the GC $\ga$-ray excess, a distinctive bump around a few GeV, 
can be well explained by DM annihilations into the SM fermions, which then 
fragment into photons.
The continuous photon spectrum mainly comes from the decay of neutral pions, which are originated
from the fragmentation of the quarks in the annihilation of the dark matter. In addition, the quarks can also
fragment into charged pions, which subsequently decay into muons and eventually electrons. The dark matter
can also directly annihilate into taus, muons, and electrons. The taus and muons will eventually decay into
electrons. Although all these electrons undergo the inverse Compton scattering and bremsstrahlung, 
which can only give rise to photons at the lower photon energy, 
it does not effect the region of $E_\gamma>1\gev$~\cite{Cumberbatch:2010ii}. 
As a result, we do not consider inverse Compton scattering and bremsstrahlung in this study. 

\subsection{DM relic abundance} \label{sec:DM_relic}

The DM relic density can be obtained by solving the Boltzmann equation for 
the DM
density evolution with the thermally-averaged annihilation cross-section into 
SM fermions. 
In this work, we assume that the thermal relic scenario such that 
the current relic density is determined by the DM annihilation 
and coannihilation of the excited state, and the number densities of 
these particles follow the Boltzmann distribution
before freeze-out.
Note that, in the context of the relic density calculation, one cannot simply assume 
$v \ll 1$, that is only valid
in the $X$-ray and $\gamma$-ray flux computation. Instead,
one has to properly take into account the thermal average effect. 
Following Ref.~\cite{Griest:1990kh},
we compute the thermal relic density from the thermally averaged 
annihilation cross-section based on
Eqs.~\eqref{eq:x1x1_ff}, \eqref{eq:p2p2_ff_sim} and \eqref{eq:p1p1_ff}. 
However, the effective relativistic degrees of freedom are
taken from the default numerical table of DarkSUSY~ \cite{Gondolo:2004sc}. 
Also, we use the PLANCK result of $\Omega h^2=0.120$~\cite{Ade:2013zuv} together with the
$10\%$ theoretical error to constrain the relic density.

A comment on the DM density computation is in order here.  Due
to a small mass splitting of 3.5 keV between the DM particle and
excited state to account for the X-ray line, coannihilation processes
involving the excited state have to be taken into account. As
mentioned above,  we focus on the scenario with the
resonance enhancement via the $X^3$ exchange. As a result, the only
relevant interactions are the DM annihilation and excited state
annihilation into SM fermions. For the Majorana DM case, the dominant
contribution to relic abundance comes from the excited state 
annihilation, $\psi_2 \bar{\psi}_2 \to \bar{f} f$, which is dominated by
$S$-wave due to the Dirac nature 
of $\psi_2$, while $\chi_1 \chi_1 \to \bar{f} f$ is $P$-wave
suppressed because of $\chi_1$ being Majorana.  
Furthermore, the large resonance enhancement 
in the process $\chi_1\chi_1\to \bar{f} f$ required to explain the $X$-ray
line at current time ($v \sim 10^{-3} c$) 
is no longer the case at the time of freeze-out, 
because during the freeze-out the relative velocity is much larger of order 
$\sim \frac{1}{3} \, c$ such that the annihilation deviates considerably 
away from the resonance region.
Therefore, $\sigmav(\chi_1\chi_1\to \bar{f} f)$ at freeze-out
is much smaller than the current annihilation cross-section $10^{-26}$
cm$^3$sec$^{-1}$, which is the right size to accommodate 
the GC $\ga$-ray excess.\footnote{One might think that $S$-wave dominated interactions
  will have the same cross-section at freeze-out as the current one
  while $P$-wave dominated ones have the larger cross-section at
  freeze-out due to the larger DM velocity~($\sim\frac{1}{3}\, c$)
  compared to the current velocity~($\sim 10^{-3}$\, c). However, it is
  not always true, especially when the resonance enhancement takes
  place as we shall see later.  
} 
Hence, $\chi_1\chi_1\to \bar{f} f$
alone cannot give rise to the correct relic density, which roughly
requires an annihilation cross section of $3\times10^{-26}$ cm$^3$sec$^{-1}$. 
For the Majorana DM case, this problem can be circumvented by 
the $S$-wave process $\psi_2 \bar{\psi}_2 \to \bar{f} f$, which
can give an annihilation cross section of order $10^{-26}$ cm$^3$sec$^{-1}$
to explain the relic abundance.

In the Dirac DM case, however, both $\psi_1$ and $\psi_2$ are Dirac
particles and all processes are $S$-wave dominated. In the context of
the DM density, annihilations of $\psi_1$ and those of $\psi_2$
contribute almost equally due to the nearly degenerate mass spectrum. 
However, the annihilation cross sections of both processes
are much smaller than $10^{-26}$ cm$^3$sec$^{-1}$ at
freeze-out due to the deviation from the resonance region as explained above.
Therefore, one has to involve 
an additional DM annihilation mechanism
to reduce the relic abundance.  A possible solution, for instance, is
to embed $(\chi_2, \chi_1)^T$ into a larger multiplet such as
$(\chi_4,\chi_2, \chi_1,\chi_3)^T$ such that 
coannihilations between $\chi_1$ and $\chi_3$ via $X^{1,2}$ is possible 
to bring
down the relic density.  As long as the mass difference between $m_{\chi_3}$ and
$m_{\chi_1}$ is much larger than 3.5 keV, $\chi_3$ cannot be generated
currently and thus the existence of $\chi_3$ is irrelevant to the
X-ray line and $\ga$-ray excess.

We summarize the discussion here with Table~\ref{tab:comp_all}, 
where we show the cross-sections in orders of magnitude 
at the time of freeze-out and the current time. 
It is clear that only Majorana DM can accommodate the correct 
relic density due to the dominant  contribution from the S-wave 
process $\bar{\psi}_2 \psi_2 \to \bar{f}f$.
\begin{table}[!h!]
\centering
\begin{tabular}{ c c c c}
  \hline\hline
  Majorana DM & $\lan \sig v \ran_{\chi_1\chi_1\to \bar{f} f}$ &  $\lan \sig v \ran_{\bar{\psi}_2\psi_2\to \bar{f} f}$ 
   &  $\lan \sig v \ran_{\chi_1\chi_1\to \bar{\psi}_2 \psi_2} $     \\
  \hline
 freeze-out & $10^{-31}$ & $10^{-26}$ & $10^{-24}$  \\
 \hline
  current & $10^{-26}$~(GC $\ga$-ray) & 0 & $10^{-19}$~(X-ray)  \\
 \hline\hline
  Dirac DM & $\lan \sig v \ran_{\bar{\psi}_1\psi_1\to \bar{f} f}$ &  $\lan \sig v \ran_{\bar{\psi}_2\psi_2\to \bar{f} f}$ 
   &  $\lan \sig v \ran_{\bar{\psi}_1\psi_1\to \bar{\psi}_2 \psi_2} $     \\
  \hline
 freeze-out & $10^{-31}$ & $10^{-31}$ & $10^{-24}$  \\
 \hline
  current & $10^{-26}$~(GC $\ga$-ray) & 0 & $10^{-19}$~(X-ray)   \\
 \hline\hline
\end{tabular}
\caption{\emph{Relevant cross-sections in orders of magnitude
in units of cm$^{-3}$sec$^{-1}$ 
at the freeze-out and at the current time. $\langle \sigma v \rangle_{\chi_1\chi_1 \to \bar{\psi}_2 \psi_2}$
and $\langle \sigma v \rangle_{\bar{\psi}_1\psi_1 \to \bar{\psi}_2 \psi_2}$
at freeze-out are not relevant for DM relic density computation.  }}
\label{tab:comp_all}
\end{table}

\section{Results }\label{sec:data analysis}

In order to employ the resonance enhancement, we rewrite 
$m_{X^3}=(2-\delta) m_{DM}$ with $m_{DM}=m_{\chi_1}~(m_{\psi_1}$)
in the Majorana~(Dirac) DM case. Therefore, at phenomenological level
 we choose $m_{\chi_1}$, $g_X$, $\delta$ and $\sc$ as 4 independent
input parameters to investigate if the proposed non-abelian DM models can 
simultaneously account for the GC $\gamma$-ray excess and 
the 3.5 keV X-ray line, and
thermally reproduce the correct relic abundance.

\begin{figure}[!htb!]
  \begin{minipage}{0.45\textwidth}
   \includegraphics[scale=0.40]{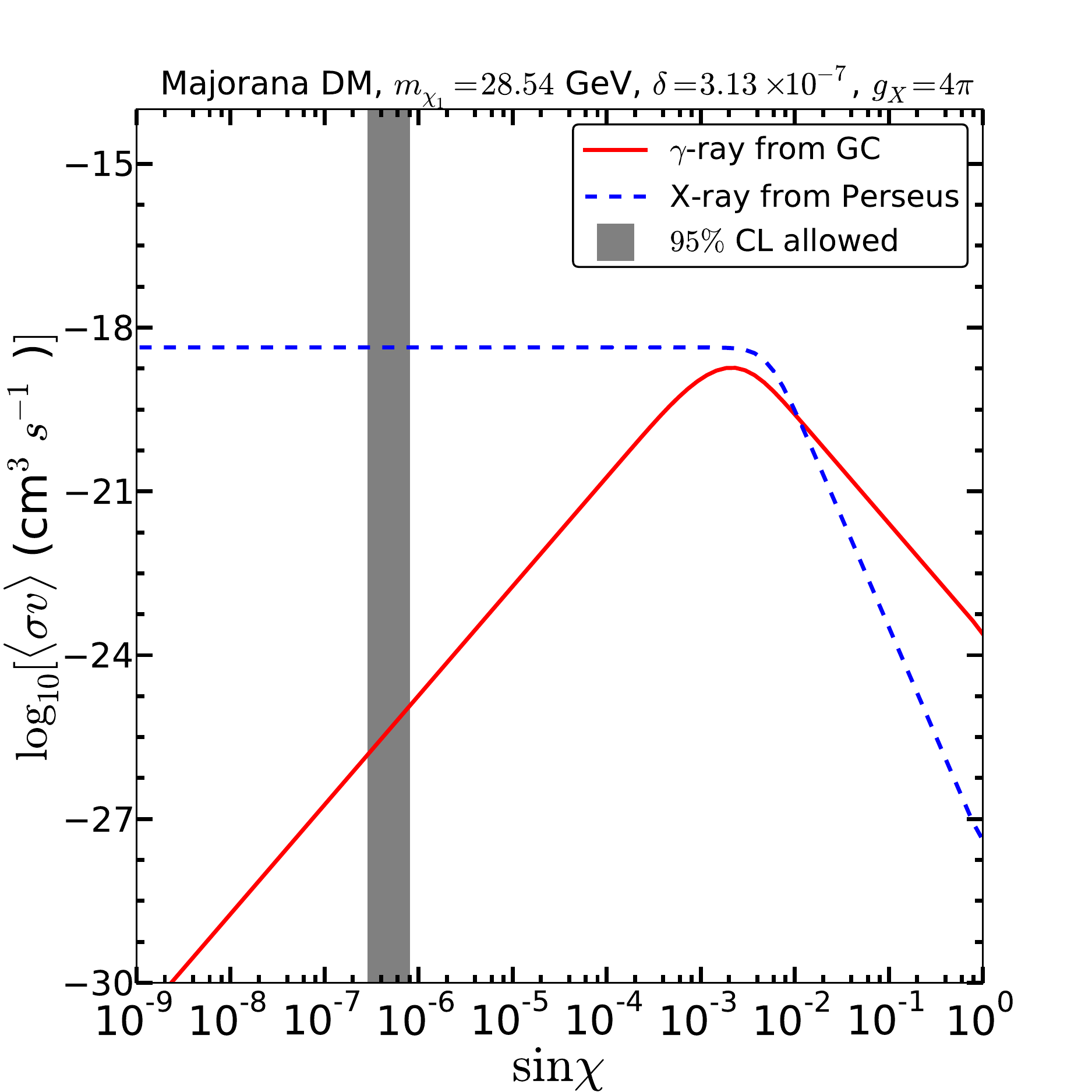}
    \end{minipage}\hfill
  \begin{minipage}{0.45\textwidth}
   \includegraphics[scale=0.40]{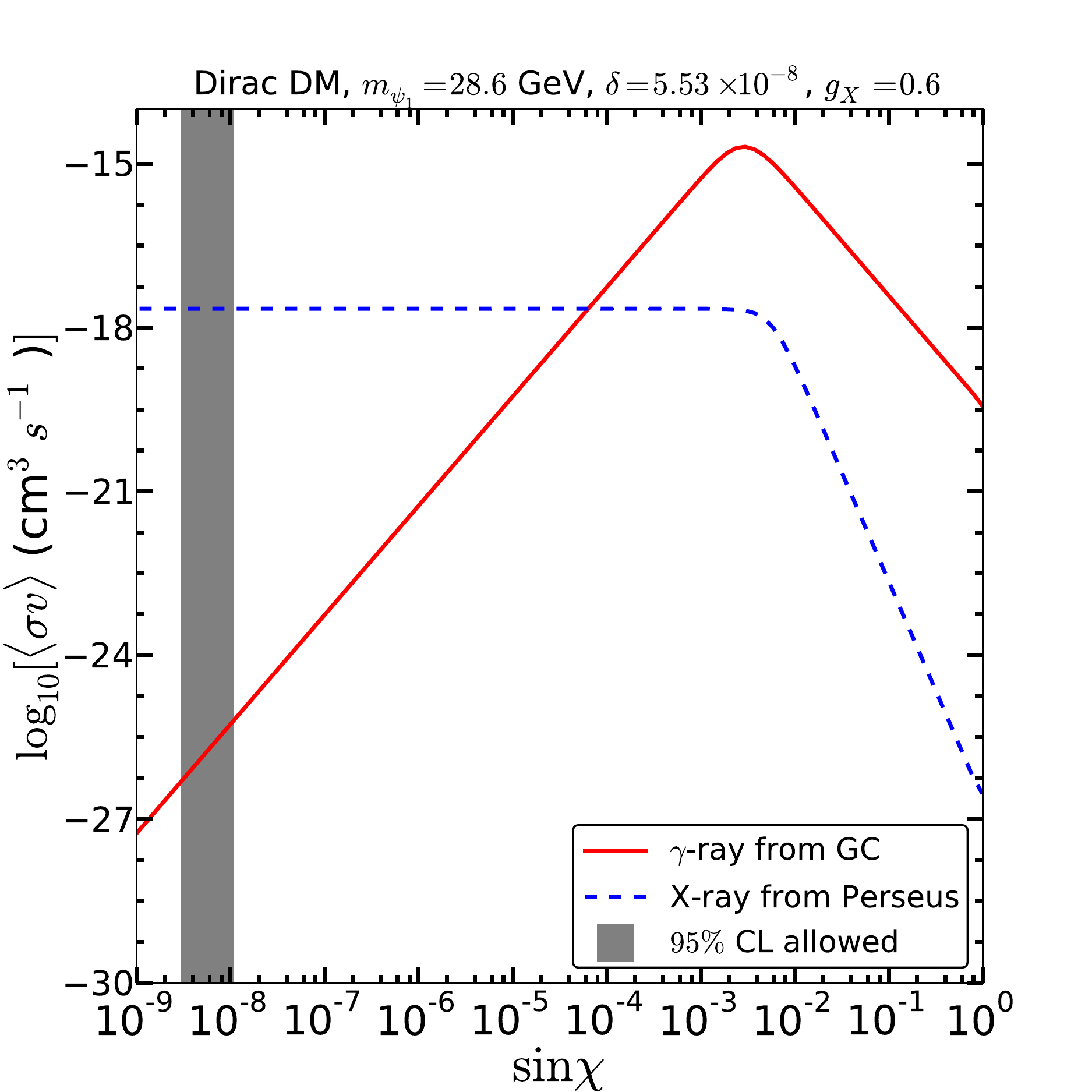}
    \end{minipage}\hfill
     \caption{\emph{Left panel: the Majorana case. Right panel: the Dirac case. The grey band represents the $95\%$ confidence region for
     the $\gamma$-rays and X-rays data.}}
\label{fig:sinx_sv}
\end{figure}

Before moving into the numerical analysis, we would like to comment on the 
region of interest for $\sc$.      
For illustration, we choose the Majorana DM case but the Dirac DM case 
exhibits the same feature. 
As mentioned above, we aim for $\lan \sigma v \ran_{\chi_1\chi_1 \to 
\bar{\psi}_2 \psi_2} \sim 10^{-19}$ cm$^3$ sec$^{-1}$
to explain the 3.5 keV X-ray line and $\lan \sigma v \ran_{\chi_1\chi_1 \to 
\bar{f} f} \sim 10^{-26}$ cm$^3$ sec$^{-1}$ to realize
the GC $\gamma$-ray excess. On the other hand, we have from 
Eq.~(\ref{eq:x1x1_ff_sim}) and 
~(\ref{eq:x1x1_x2x2_sim})
in the limit of $m_{X^3}\simeq 2 m_{\chi_1}$,  $v \ll 1$ and $m_f \simeq 0$,
\bea
\lee \sigma v \rii _{\chi_1\chi_1 \to \bar{f} f} &\sim&
\frac{\sqrt{s}}{ \lee \lee s -m^2_{X^3} \rii^2 + \Ga^2_{X^3} m^2_{X^3} \rii}  \lee g^2_L + g^2_R\rii v^2
\sim \frac{\sqrt{s} }{ m^4 v^4_{vrel}  + 4 m^2_{\chi_1} \Ga^2_{X^3}  } \sin^2\chi \, v^2 ,
\nn\\
\lee \sigma v \rii _{\chi_1\chi_1 \to \bar{\psi}_2 \psi_2} &\sim&
\frac{ \sqrt{s -4 m^2_{\psi_2}}  }{ \lee \lee s -m^2_{X^3} \rii^2 + \Ga^2_{X^3} m^2_{X^3} \rii} v^2
\sim \frac{ \sqrt{s -4 m^2_{\psi_2}} }{ m^4 v^4_{vrel}  + 4 m^2_{\chi_1} \Ga^2_{X^3} }   v^2 ,
\eea
where we have suppressed the kinematics factors and the coupling constant $g_X$, which do not 
affect the argument.
It is clear that in order to achieve $\svxr/\lan \sigma v\ran_{\ga} \sim 10^7$,
one must have $\sin^2\chi$ smaller than $10^{-7}$.
It implies that in the denominator of the cross-section, 
$\Ga^2_{X^3} m^2_{\chi_1}\sim \lee \frac{1}{16\pi^2 } \rii^2 m^4_{\chi_1}
 \sin^4\chi$ becomes
negligible compared to $m^4_{\chi_1} v^4$ with $v\sim 10^{-3}\,c$. In 
other words, we saturate the resonance enhancement since the DM
velocity becomes dominant in the denominator and 
any further decrease in $\sc$ will not affect the cross-section.
In Fig.~{\ref{fig:sinx_sv}}, we can clearly see that for both the Majorana and Dirac 
case,  $\sc$ is located in the saturated area, i.e., $\sc \ll v$.
Furthermore, the reasons why the required mixing is so small, 
$\sin^2\chi \ll 10^{-7}$, are because
first, 
$\lan \sig v \ran_{\chi_1\chi_1 \to \bar{\psi}_2 \psi_2}$
has a large kinematical suppression factor, $\sqrt{s - 4 m^2_{\psi_2}}$, 
compared to $\lan \sigma v \ran_{\chi_1\chi_1 \to \bar{f} f}$
and second, the $v$ for X-rays in the Perseus cluster is larger than the $v$ for $\ga$-rays 
in the GC. 
Subsequently, $\lan \sig v \ran_{\chi_1\chi_1 \to \bar{\psi}_2 \psi_2}$
is much smaller than $\lan \sigma v \ran_{\chi_1\chi_1 \to \bar{f} f}$ with 
$\sin\chi \sim 1$. It 
indicates that one actually needs $\sin^2\chi \ll 10^{-7}$
in order to fulfill $\svxr /\lan \sigma v\ran_{\ga} \sim 10^7$.
    
For the fitting procedure, we make use of the minimum chi-squared method. 
Since the likelihoods for the relic density, X-ray line, and GC $\gamma$-rays data 
are well Gaussian-distributed, and the $95\%$ and $99.73\%$ confidence 
limits  in two-dimensional contour plots 
correspond to $\delta\chi^2=5.99$ and $\delta\chi^2=11.83$, respectively.

\subsection{Majorana case}\label{sec:data Majorana}

\begin{figure}[!htb!]
  \begin{minipage}{0.45\textwidth}
   \includegraphics[scale=0.45]{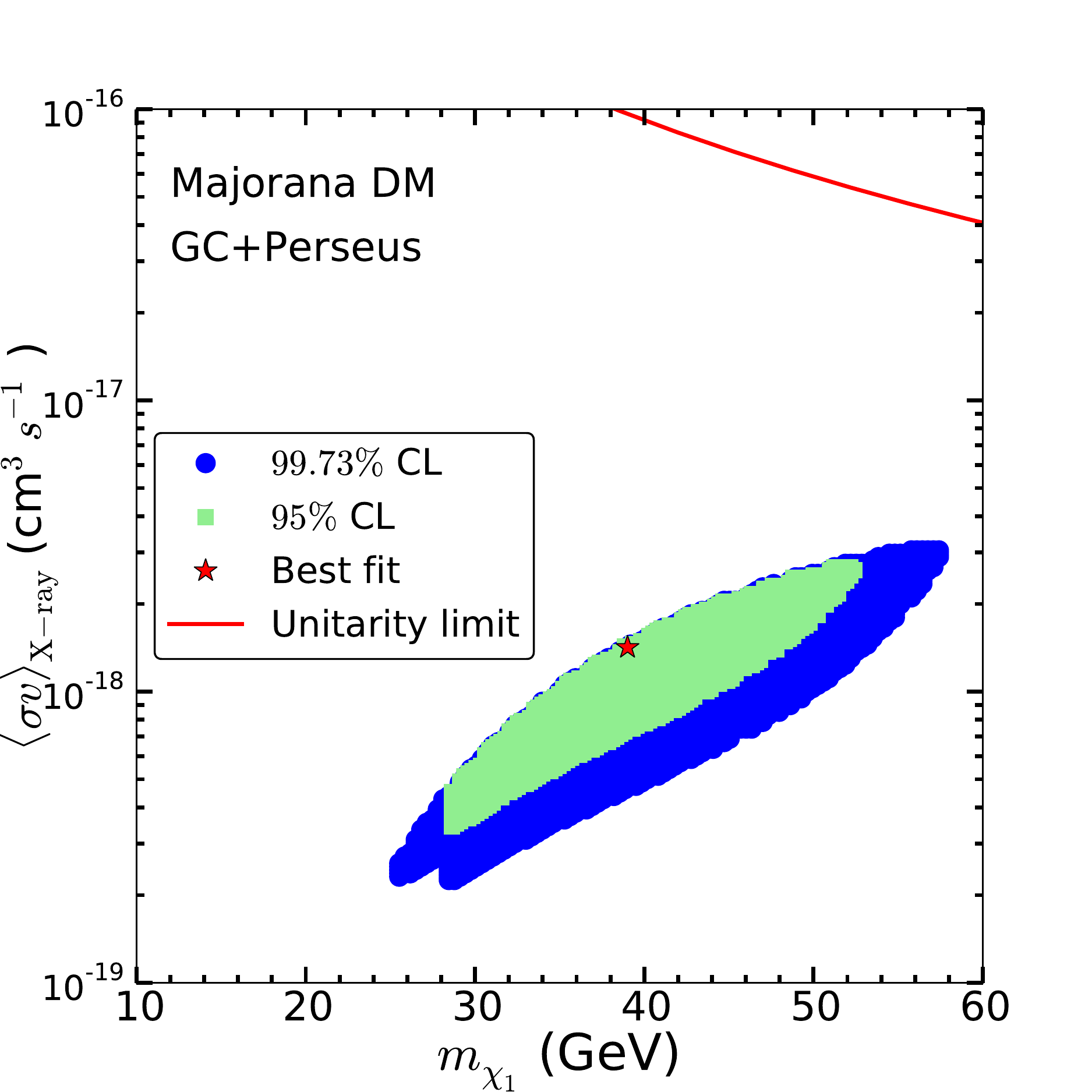}
    \end{minipage}\hfill
  \begin{minipage}{0.45\textwidth}
   \includegraphics[scale=0.45]{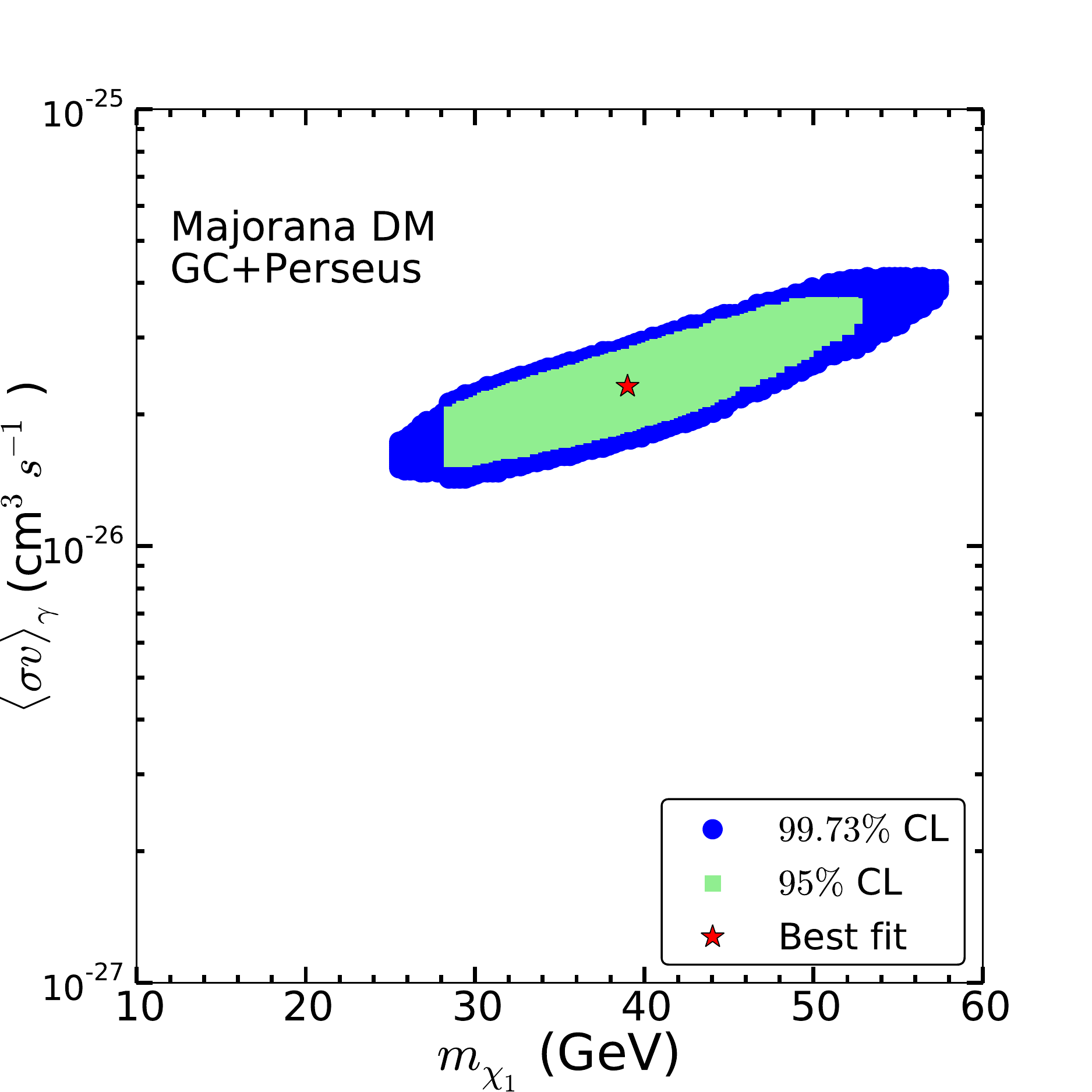}
    \end{minipage}\hfill 
     \begin{minipage}{0.25\textwidth}
   \includegraphics[scale=0.45]{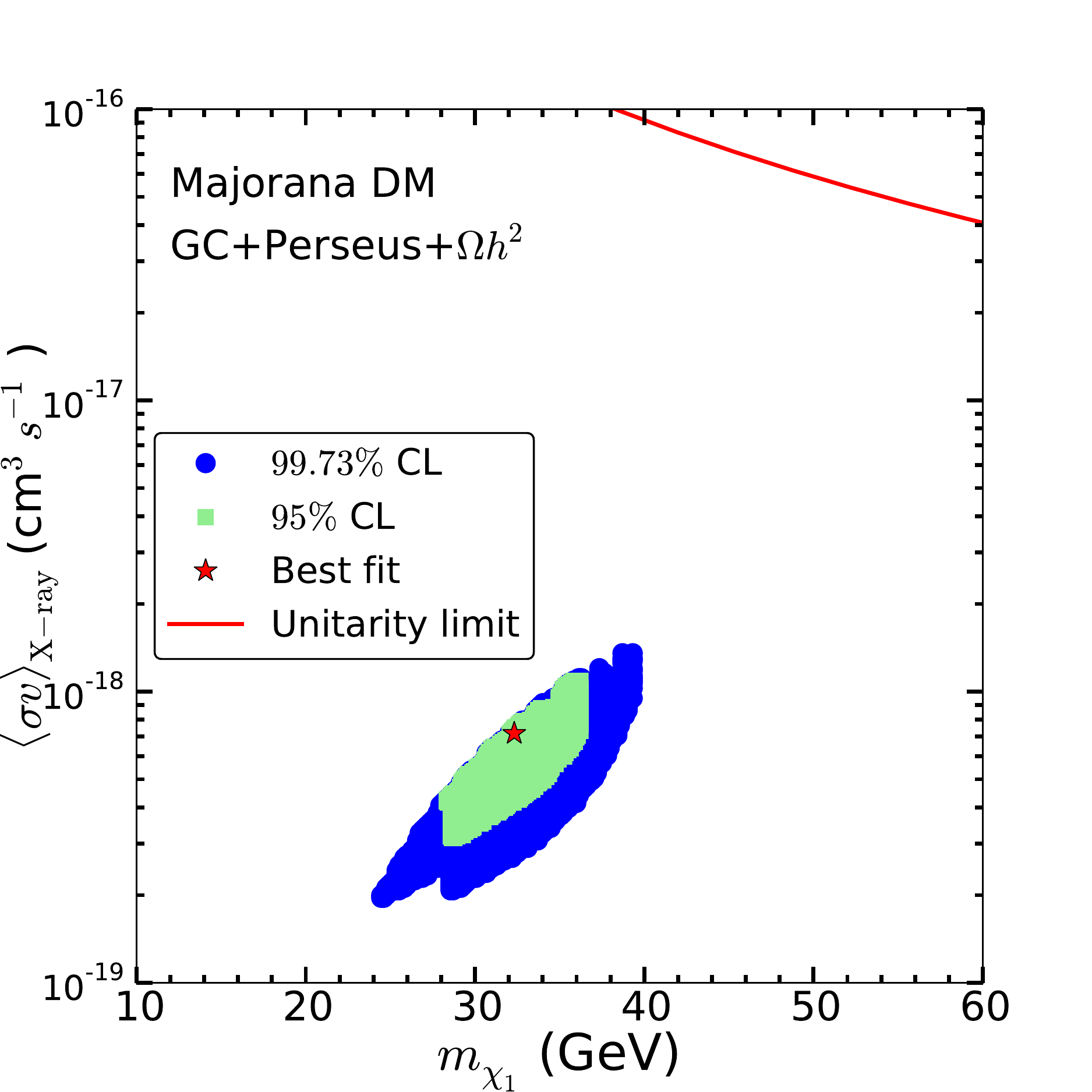}
    \end{minipage}\hfill
  \begin{minipage}{0.45\textwidth}
   \includegraphics[scale=0.45]{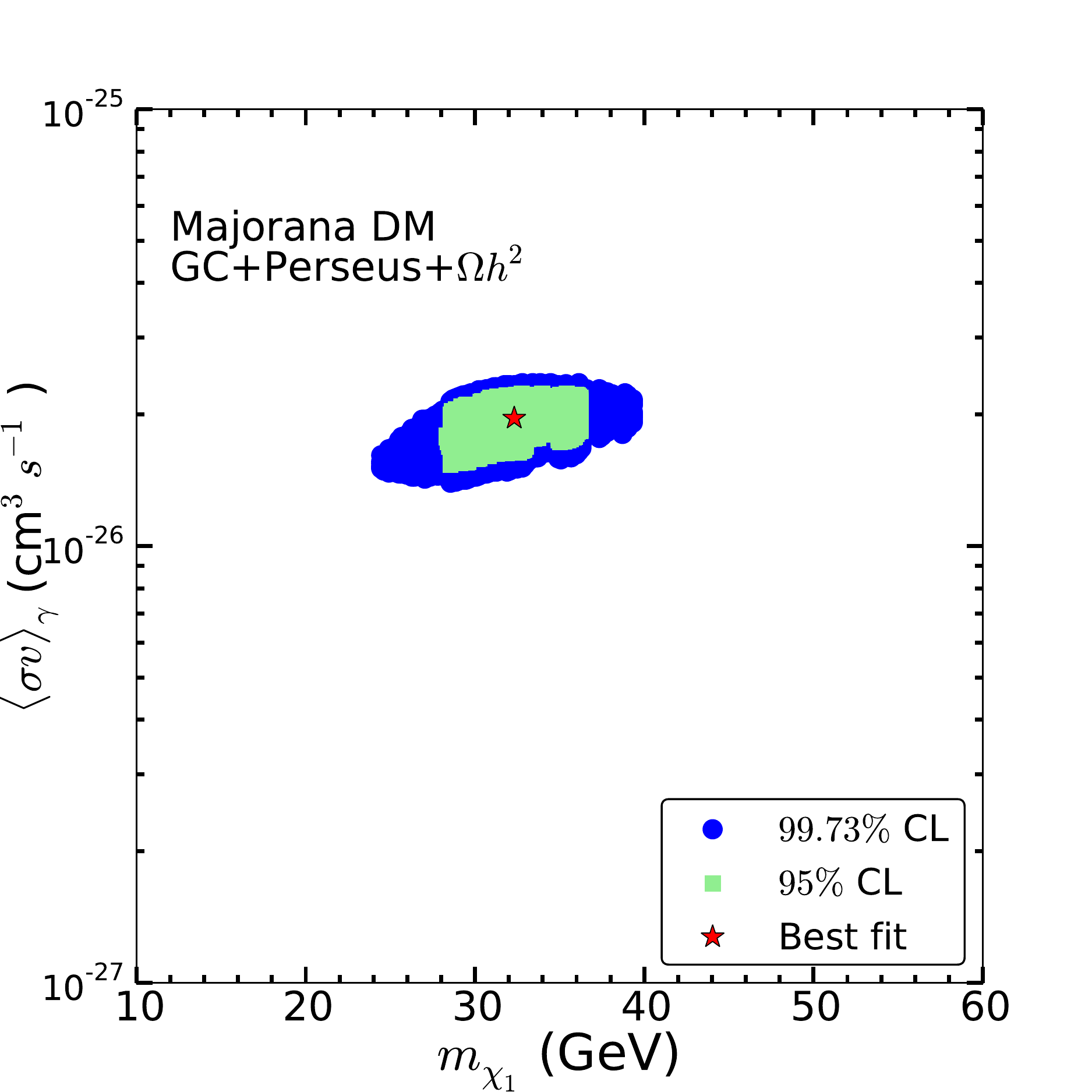}
    \end{minipage}\hfill
     \caption{\emph{The $95\%$ and $99.73\%$ confidence-level regions
in the plane of annihilation cross sections versus the DM mass
 obtained in the fits with (i) GC gamma-ray and Perseus X-ray data 
(upper panels) and (ii) also the DM relic density (lower panels) 
for the Majorana DM case.
We show on the left panels: 
   $\langle \sigma v \rangle_{\rm X-ray}$ versus $m_{\chi_1}$ for X-rays,
and on the right panels: $\langle \sigma v \rangle_{\gamma}$ 
versus $m_{\chi_1}$ for $\gamma$-rays.}}
\label{fig:mx_sv_Maj}
\end{figure}

In this section, we present the results in the Majorana DM case with
the Majorana DM $\chi_1$ and the Dirac excited state $\psi_2$.
Throughout this (and also
next) section, the way we present the results is to project confidence
regions into planes of parameters or observables.\footnote{In Fig.~\ref{fig:mx_sv_Maj},
for instance, we project the confidence
region into the $\svxr$-$m_{\chi_1}$~(left panels) and $\lan \sig
  v\ran_{\ga} $-$m_{\chi_1}$~(right panels) plane.} 
In the figures, inside the legend: ``GC+Perseus'' means that 
the confidence regions are obtained from the fit
with only the GC $\ga$-ray and X-ray data, while
``GC+Perseus+$\Omega h^2$'' indicates that the DM relic density is also
included in the fitting in addition to the $\ga$- and
X-ray data.

In Fig.~\ref{fig:mx_sv_Maj}, we show the confidence regions in terms of 
the DM annihilation cross-section for the X-ray and $\ga$-ray
versus the DM mass.
In the upper panels, we include the GC $\ga$-ray excess and X-ray line only
while the DM density is also included in the lower panels. 
The green (blue) area corresponds to the $95\%$~$(99.73\%)$ 
confidence region while the red star represents the best-fit point.  
Furthermore, the unitarity bound~\cite{Griest:1989wd} denoted by the 
red line comes from
\be
\lan \sig v \ran \lesssim \frac{ 3 \times 10^{-22} \,\, \mbox{cm$^3$/sec} }{ m_{\chi_1}/ \lee 1 \,\, \mbox{TeV} \rii }. 
\ee
We would like to make the following comments. 
\begin{itemize}
\item The corresponding X-ray cross-section~(left panels) is centered
  around $10^{-18}$ cm$^3$/sec consistent with Eq.~(\ref{eq:Xrayflux})
  while for $\ga$-rays~(right panels), one needs $\lan \sig v \ran
  \sim 10^{-26}$ cm$^3$/sec complying with Ref.~\cite{Calore:2014xka}.

\item The $\ga$-ray spectrum coming from 
$\chi_1 \chi_1 \to \bar f f$ certainly depends on the final
  states. In this model, the final states include both quarks and
  leptons, and the final state  composition is fixed according to
  Eq.~(\ref{eq:Xsec_xray}) and (\ref{eq:gLgR}). In general, a quark final
  state demands a higher DM mass due to soft photon spectra compared to
  a leptonic one. Therefore, $m_{\chi_1}$ will lie between that of
  the purely $b$-quark case and that of purely $\tau$ case.

\item For X-ray plots~(left panels), there exists a sharp cut-off
  close to the best-fit point on the top of the confidence
  regions. It is due to the perturbative limit: $g_X \leq 4\pi$ as we
  shall see later the best-fit point has $g_X$ quite close to
  $4\pi$. In contrast, the best-fit points for $\ga$-ray plots~(right panels)
  are located near the central area of the confidence
  region, which comes from the fact the $\lan \sig v\ran_{\ga} $ can
  be enlarged by increasing the mixing $\sc$ between the SM and dark
  sector without varying $g_X$ as shown in
  Fig.~\ref{fig:sinx_sv} while $\svxr$ is insensitive to $\sc$ in the
  saturated area.

\item All plots exhibit a sharp cut-off on $2\sig$ regions especially
  on the left-hand side. It is due to the fact that GC $\ga$-ray bump
  shown in Fig.~\ref{fig:GC_calore} has a sharp drop around 0.5 GeV
  compared to the milder change on the right hand side around 20
  GeV.  Consequently, the bump of the predicted photon spectrum
  will not coincide with that of the GC excess, leading to a surge in
  chi-square, once $m_{\chi_1}$ becomes much smaller than the best-fit
  value.

\item As explained in Section~\ref{sec:DM_relic}, the Majorana DM
 case can accommodate the correct DM density with the $S$-wave process
$\psi_2 \psi_2 \to \bar{f} f$ being the main contribution. 
 Including the DM relic constraint reduces the confidence region 
significantly; only in the region of $ 25 \lesssim m_{\chi_1} \lesssim 40$ GeV
can the model yield the correct DM density.

\end{itemize}

\begin{figure}[!htb!]
  \begin{minipage}{0.45\textwidth}
   \includegraphics[scale=0.45]{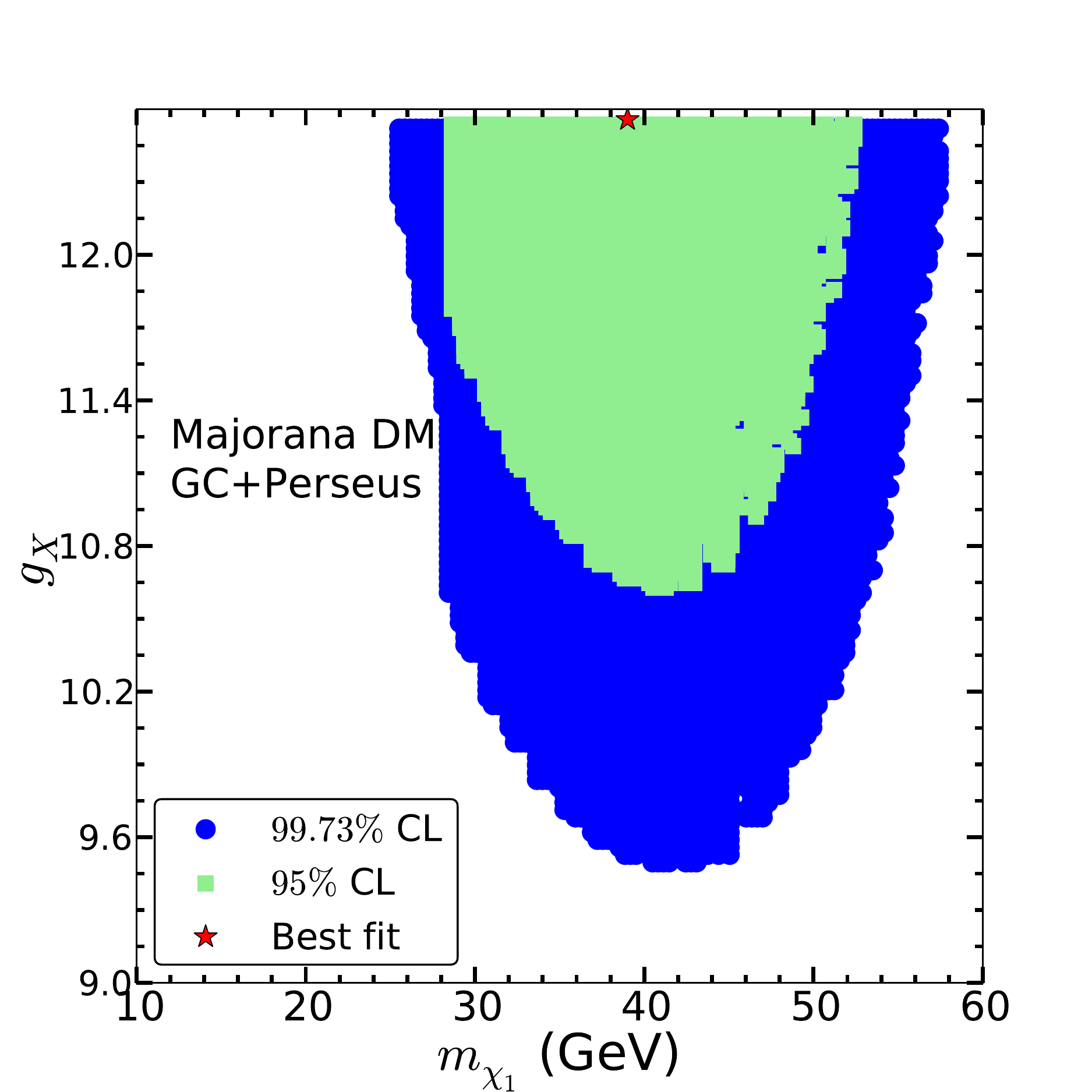}
    \end{minipage}\hfill
  \begin{minipage}{0.45\textwidth}
   \includegraphics[scale=0.45]{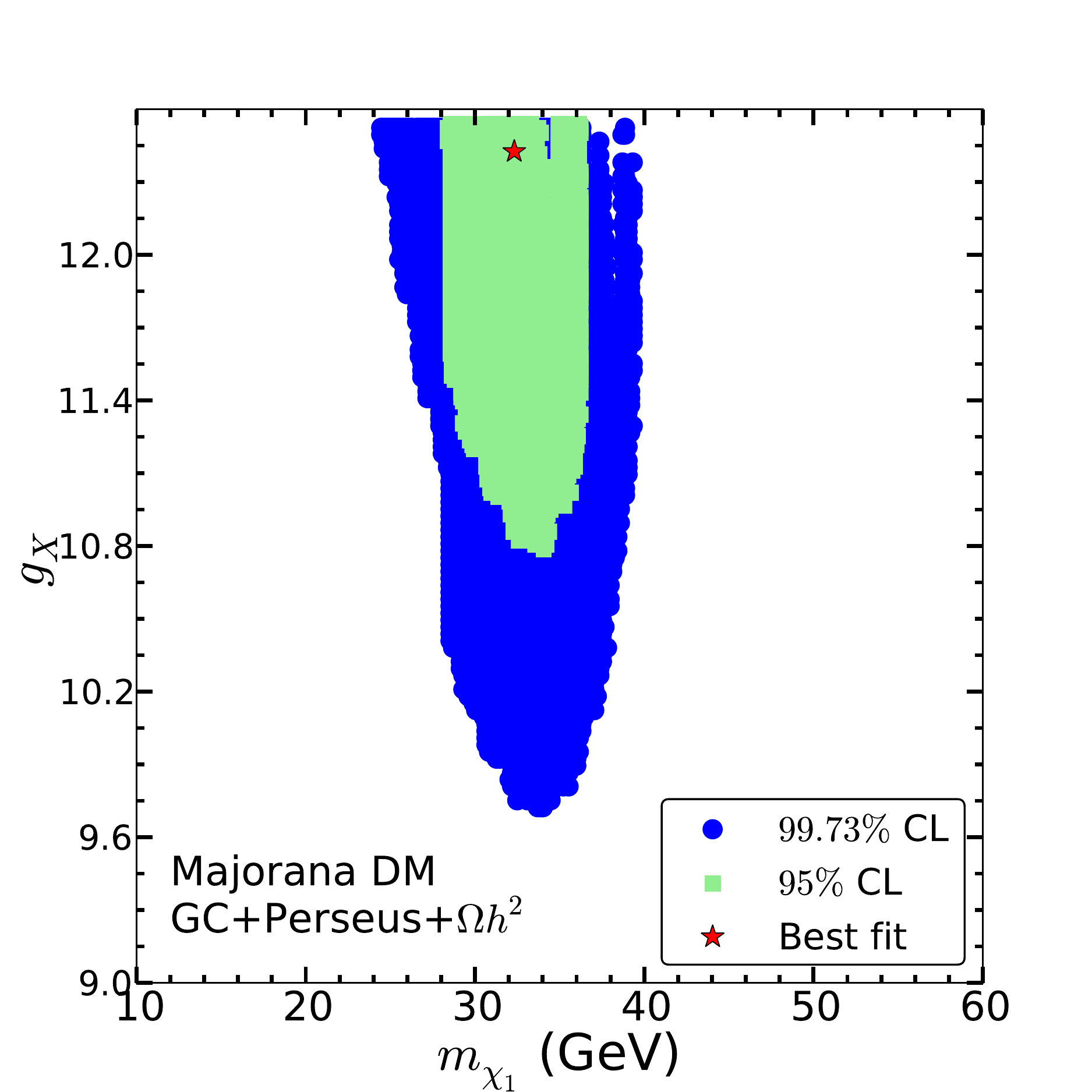}
    \end{minipage}\hfill 
     \begin{minipage}{0.45\textwidth}
   \includegraphics[scale=0.45]{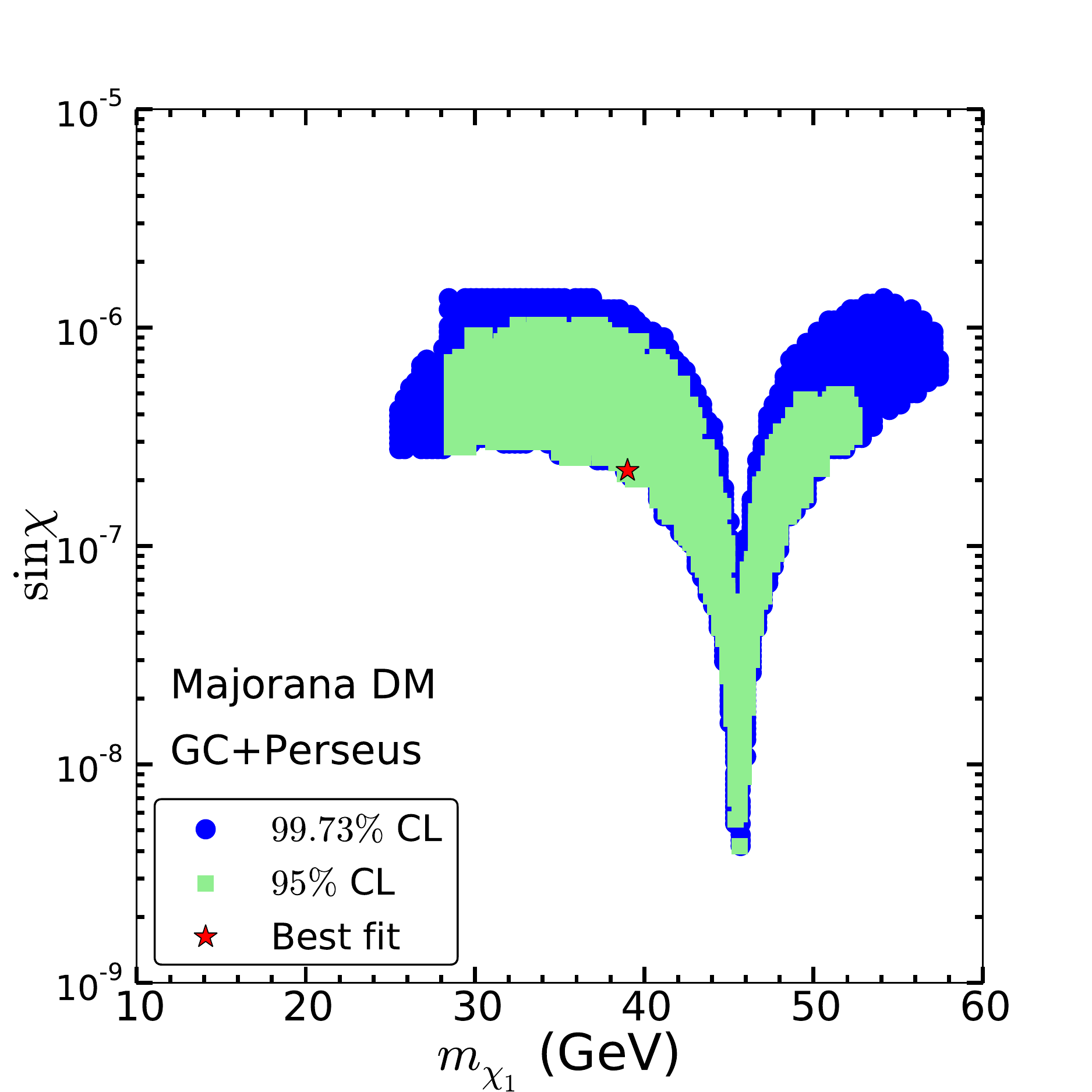}
    \end{minipage}\hfill
  \begin{minipage}{0.45\textwidth}
   \includegraphics[scale=0.45]{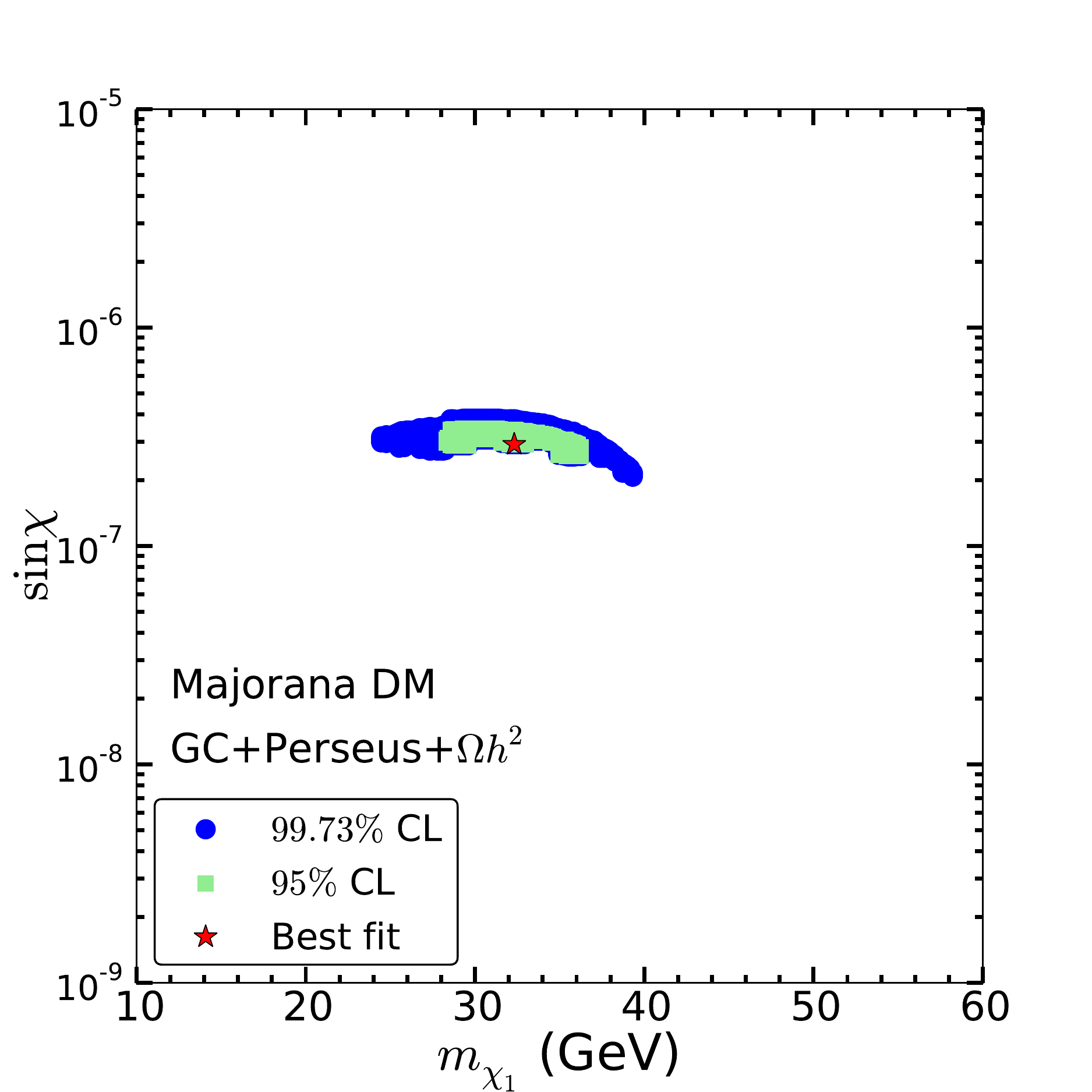}
    \end{minipage}\hfill
     \caption{\emph{The $95\%$ and $99.73\%$ confidence-level regions
in the planes of $(m_{\chi_1},\, g_X)$ (upper panels) and 
   $(m_{\chi_1},\, \sin \chi)$ (lower panels)
 obtained in the fits with (i) GC gamma-ray and Perseus X-ray data 
(left panels) and (ii) also the DM relic density (right panels) 
for the Majorana DM case.
}}
\label{fig:mx_gx_sinx_Maj}
\end{figure}

In Fig.~\ref{fig:mx_gx_sinx_Maj}, we show $g_X$ versus $m_{\chi_1}$ in
the upper panels and $\sc$ versus $m_{\chi_1}$ in the lower panels.
For the left panels, only the $\ga$-ray and $X$-ray data are included
in the fits while the DM density is also included in the right panels.
Note that both $\chi_1\chi_1 \to \bar{f} f$ responsible for the GC
$\ga$-ray excess and $\chi_1\chi_1 \to \bar{\psi}_2 \psi_2$ for the
X-ray line are P-wave suppressed by the small DM
velocity in the current Universe.
To compensate for the velocity
suppression, one needs large $g_X$ in addition to the resonance
enhancement to realize the very large $\lan \sigma v\ran$, which is
proportional to $g^4_X$, for the X-ray line. It turns out that $g_X$ is
close to the perturbativity limit $4\pi$ for the best-fit point.  In
contrast, as we shall see later, the Dirac case features $S$-wave
dominated cross-sections, i.e., without the velocity suppression, where
$g_X$ can be much smaller~($\sim 1$).  The mixing between 
the SM and the dark sector is roughly of order $10^{-7}$ but with a dip for
$m_{\chi_1}=\frac{1}{2} m_Z$. It comes from large $\zeta$ defined in
Eq.~(\ref{eq:zeta_def}) for $m_{X^3}~(=2m_{\chi_1}) \simeq m_Z$, leading
to large $g_{L,R}~(\sim \sin\zeta)$ defined in Eq.~(\ref{eq:gLgR}) and
large $\lan \sig v \ran_{\ga}$.  On the other hand, $\svxr~(\sim
\cos\zeta)$ does not change dramatically for $m_{\chi_1}=\frac{1}{2}
m_Z$.  So as to maintain $\svxr/\lan \sigma v\ran_{\ga} \sim 10^7$,
 smaller $\sin \chi$ is needed to suppress the $\ga$-ray flux with
respect to the X-ray one. Note that when $m_{X^3} \sim m_Z$, the
electroweak precision data put a stringent bound on the SM-dark
sector mixing, $\sc \lesssim 5\times10^{-3}$~\cite{Hook:2010tw}, which
is however too weak to constrain any relevant parameter space of
the model under consideration.
\begin{figure}
\vspace{-2 cm}
   \centering
   \includegraphics[scale=0.5]{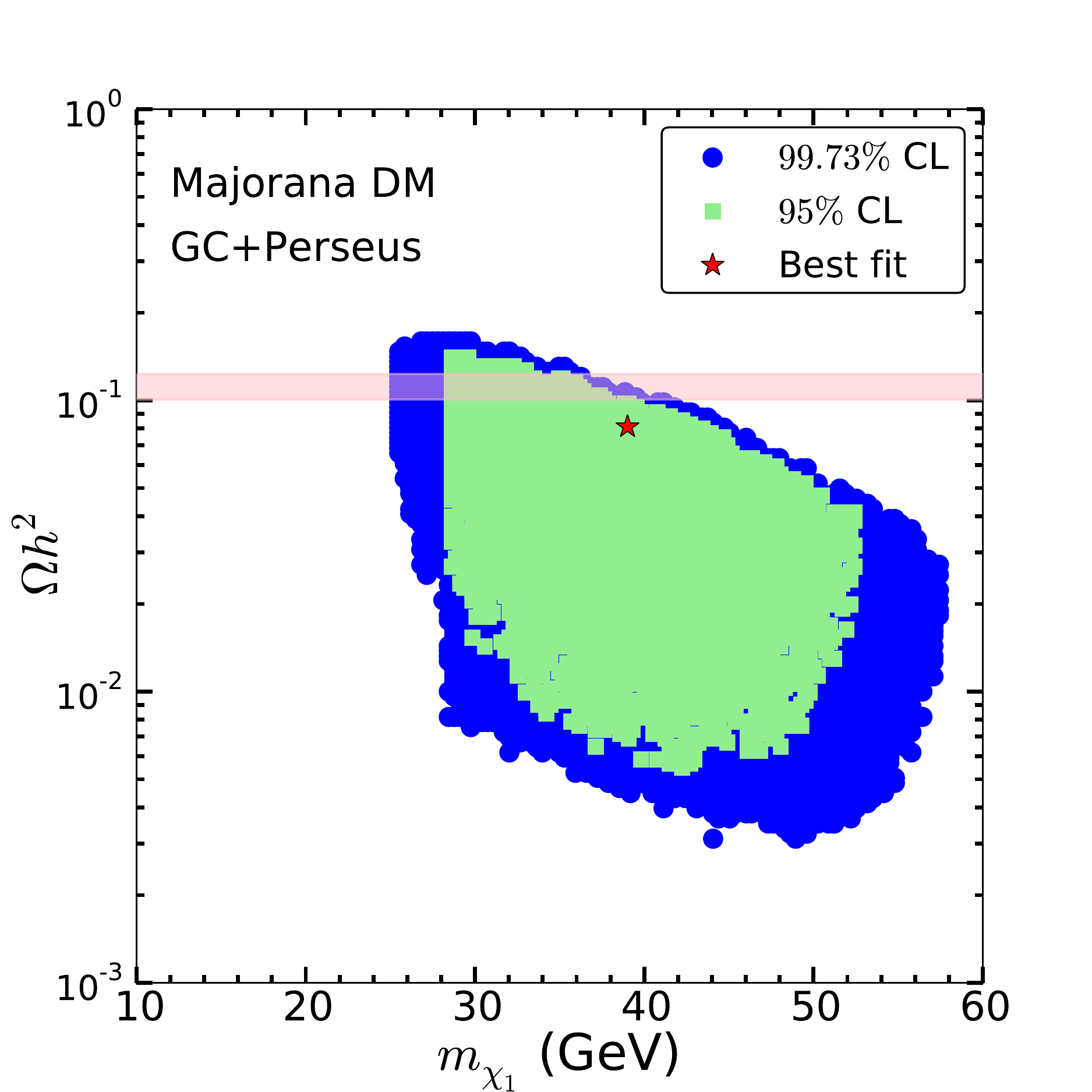}
 \caption{\emph{The $95\%$ and $99.73\%$ confidence-level regions 
in the plane of $\Omega h^2$ versus $m_{\chi_1}$ obtained in the fits with
$\ga$-ray and X-ray data only in the Majorana DM case.
The horizontal band in pink indicates 1-$\sigma$ range (including the theoretical error) of PLANCK measurement 
on the relic abundance. 
}}
 \label{fig:mx_oh2_Maj}
\end{figure}

We conclude this section with Fig.~\ref{fig:mx_oh2_Maj}, in which we
project the confidence regions, including the X-ray and $\ga$-ray data
only, into the DM relic density and $m_{\chi_1}$ plane. It is clear
that only for $25 \le m_{\chi_1} \le 40$ GeV, the correct DM density can be
reproduced.  Furthermore, the best-fit point corresponds to the
slightly lower relic density, which results in a minor shift in the
best-fit point when the DM relic density is included into the fit, as can be
seen from Fig.~\ref{fig:mx_sv_Maj} and \ref{fig:mx_gx_sinx_Maj}.
 
\subsection{Dirac case}\label{sec:data Majorana}

Here we show the results of the Dirac DM case. As argued in
Sec.~\ref{sec:DM_relic}, all relevant processes responsible for the
X-ray line, GC $\ga$-ray excess and DM relic density are $S$-wave
dominated. Moreover, the large cross-section needed to account for the
X-ray line requires the large resonance enhancement with the help of
the very small $X^3$ decay width. The large DM
velocity~($\sim\frac{1}{3}\,c$) at freeze-out implies a considerable
deviation from the resonance when DM decouples from the thermal
universe. In order to achieve $\lan \sig v \ran_{\ga}\sim 10^{-26}$
cm$^3$sec$^{-1}$ at current time, 
the cross-section of $\bar{\psi}_1\psi_1 \to \bar{f} f$ at freeze-out 
will be much smaller than $3\times10^{-26}$
cm$^3$sec$^{-1}$, the size required to reproduce the DM density. 
Therefore, we do not include the DM relic density constraint into the 
fits here.  
Notice that we have to take into account the stringent bounds on the spin-independent
DM-nucleon cross-section~\cite{Akerib:2013tjd}
due to vector-current interactions, but it hardly has any impact on the
analysis since $\sc$ of interest is extremely small, leading to a
large suppression on the DM-nucleon cross-section.

\begin{figure}[!htb!]
  \begin{minipage}{0.45\textwidth}
   \includegraphics[scale=0.45]{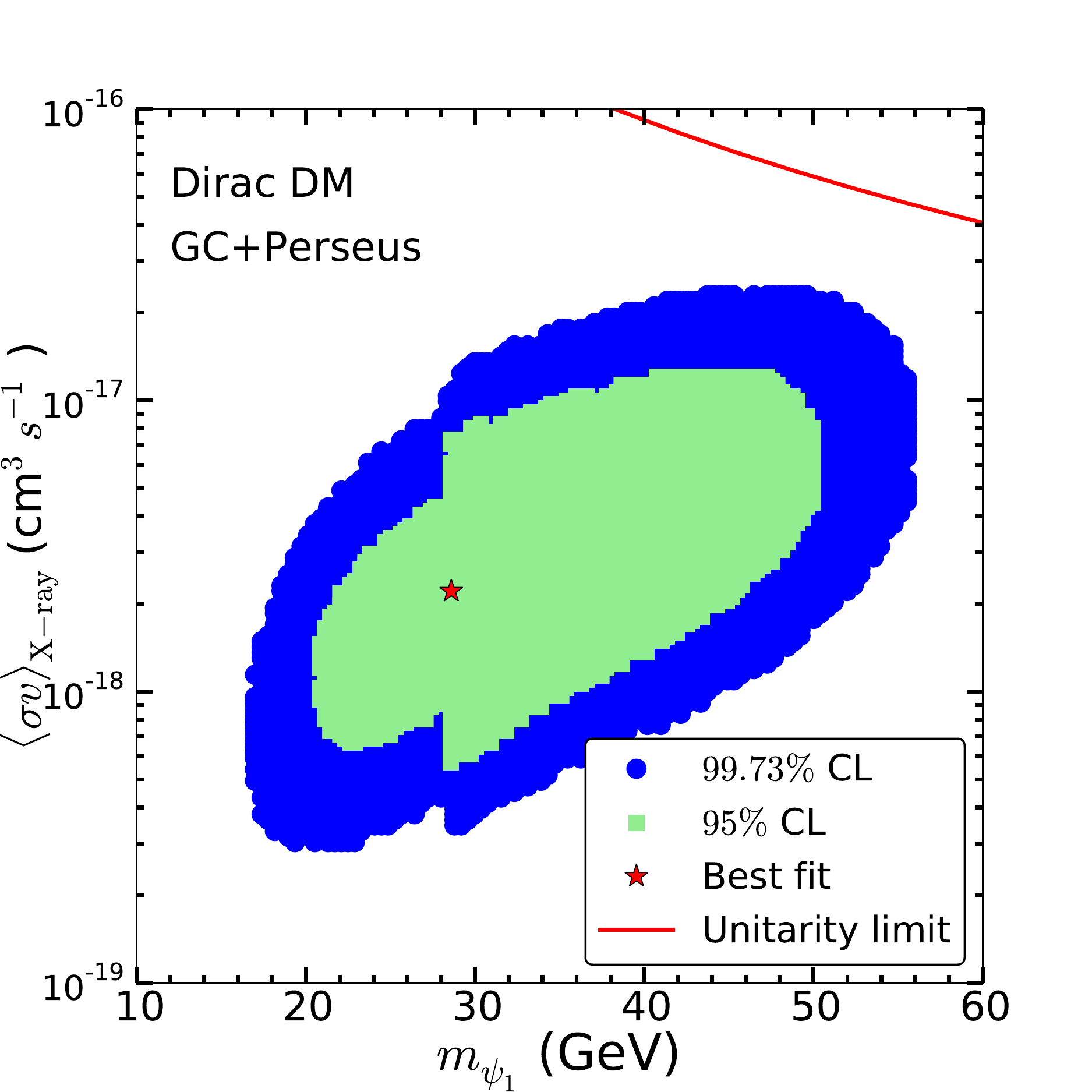}
    \end{minipage}\hfill
  \begin{minipage}{0.45\textwidth}
   \includegraphics[scale=0.45]{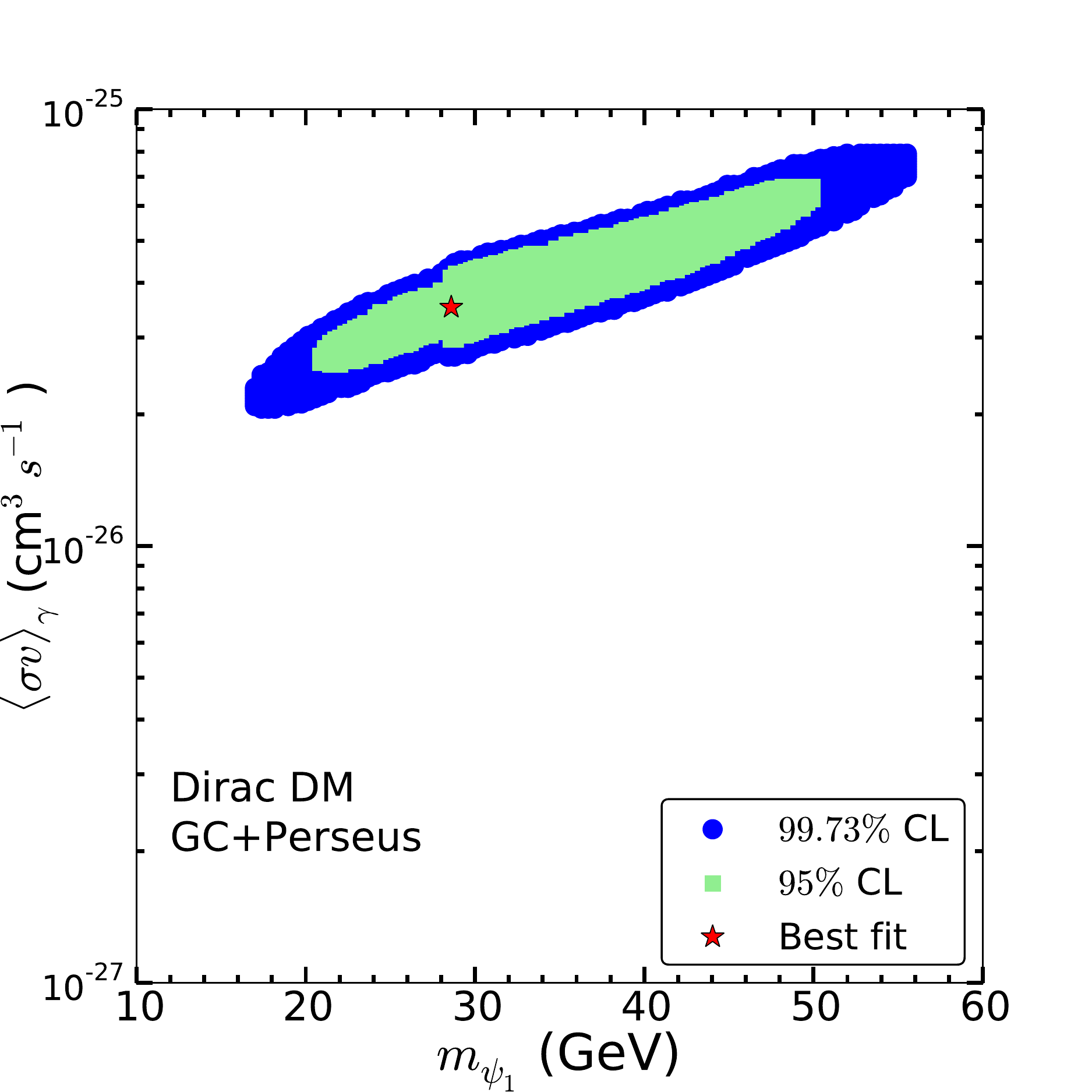}
    \end{minipage}\hfill 
     \caption{\emph{The $95\%$ and $99.73\%$ confidence-level regions 
in the plane of annihilation cross sections versus the DM mass
obtained in the fits with GC gamma-ray and Perseus X-ray data only.
for the Dirac DM case.
Left panel: $\langle \sigma v \rangle_{\rm X-ray}$ versus $m_{\psi_1}$ for 
X-rays; right panel: $\langle \sigma v \rangle_\gamma$ versus $m_{\psi_1}$ 
for $\gamma$-rays.}}
\label{fig:mx_sv_Dirac}
\end{figure}

In Fig.~\ref{fig:mx_sv_Dirac}, we show $\langle \sigma v \rangle$ 
versus $m_{\psi_1}$ for X-rays~(left panel) 
and $\ga$-rays (right panel).
Unlike the Majorana case in Fig.~\ref{fig:mx_sv_Maj}, the best-fit point in $\svxr$ is near the central area of the confidence region since
$g_X$ is much smaller than that of the Majorana case.
The steep shrink on the $2\sig$ confidence region, which is observed at the Majorana case as well, around $m_{\psi_1}\approx 29$ GeV is again
due to the sharp change on the GC $\ga$-ray spectrum around 0.5 GeV shown in Fig.~\ref{fig:GC_calore}.

\begin{figure}[!htb!]
  \begin{minipage}{0.33\textwidth}
   \includegraphics[scale=0.30]{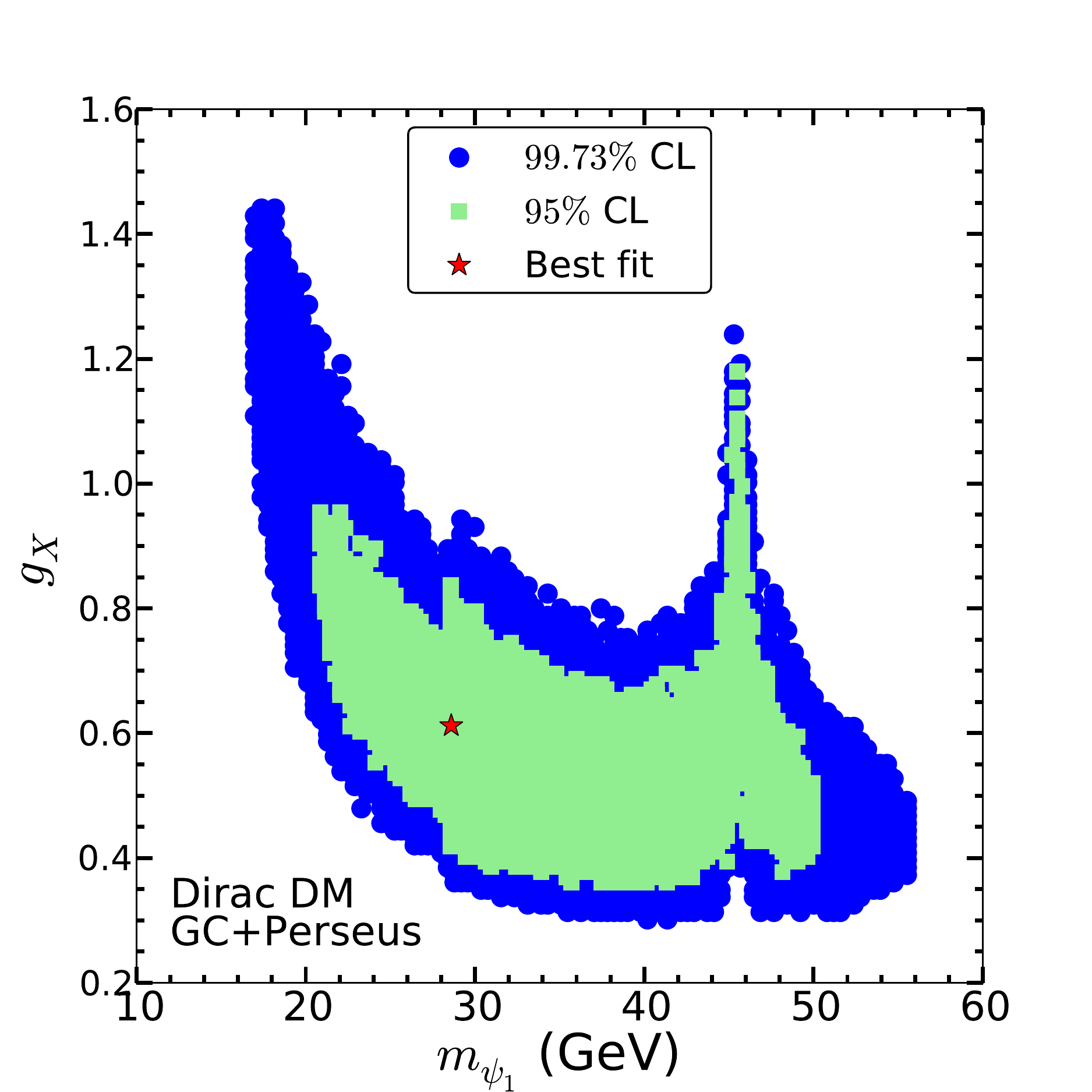}
    \end{minipage}\hfill
  \begin{minipage}{0.33\textwidth}
   \includegraphics[scale=0.30]{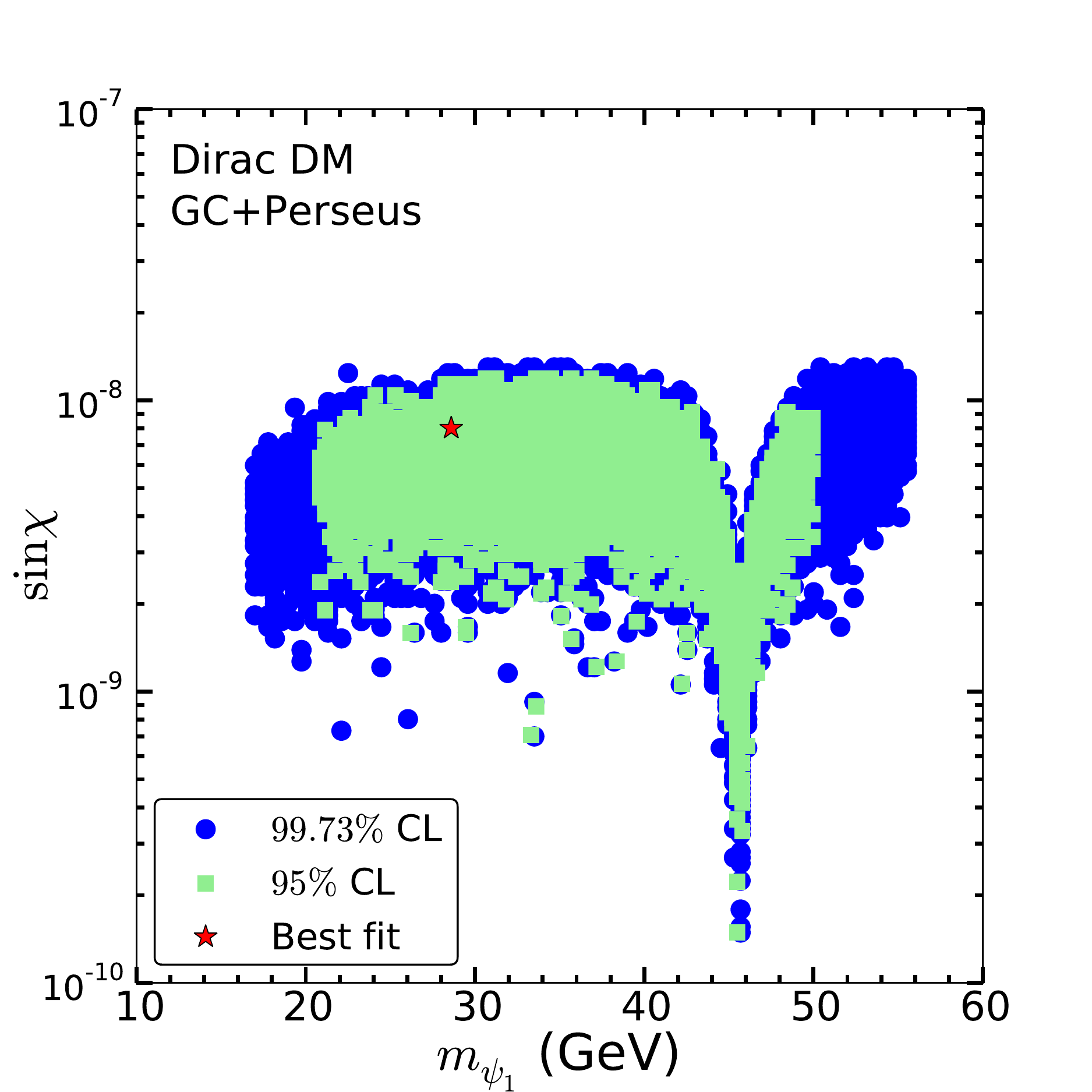}
    \end{minipage}\hfill 
      \begin{minipage}{0.33\textwidth}
   \includegraphics[scale=0.30]{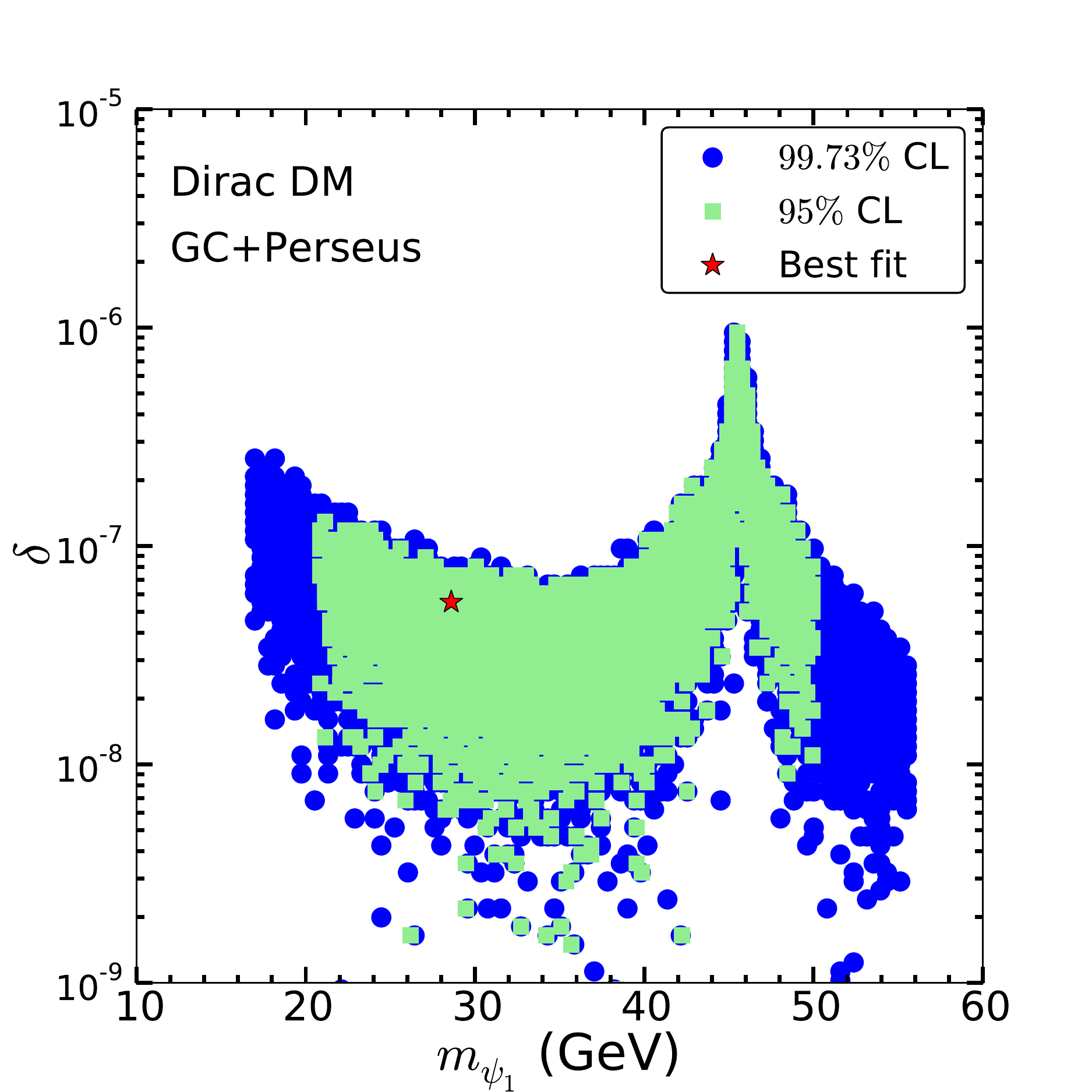}
    \end{minipage}\hfill 
     \caption{\emph{The $95\%$ and $99.73\%$ confidence-level regions
in the planes of $(m_{\chi_1},\, g_X)$ (left panel),
$(m_{\chi_1},\, \sin\chi)$ (middle), and $(m_{\chi_1},\, \delta)$ (right)
obtained in the fits with GC gamma-ray and Perseus X-ray data only
in the Dirac DM case.
}}
\label{fig:mx_gx_Dirac}
\end{figure}

In Fig.~\ref{fig:mx_gx_Dirac}, we show $g_X$, $\sc$ and $\de$ versus $m_{\psi_1}$, respectively.
All processes of interest are $S$-wave dominated
without the DM velocity suppression. It implies that the resonance
 enhancement from the narrow $X^3$ decay width alone
is sufficient to achieve the large $\svxr$ without resorting to large $g_X$. Therefore, $g_X$ is of $\mathcal{O}(0.6)$ in this case, compared
to  $g_X \sim 10$ for Majorana $\chi_1$. Similar to the Majorana DM 
case explained above, around $m_{\psi_1}~(\simeq m_{X^3})\approx \frac{1}{2} m_Z$,
$\zeta$ defined in Eq.~(\ref{eq:zeta_def}) becomes large, resulting 
in large $g_{L,R}~(\sim \sin\zeta)$ and $\lan \sig v \ran_{\ga}$.
It is then offset by the decrease in $\sc$ as shown in the left panel 
of Fig.~\ref{fig:mx_gx_Dirac}. At the same time, from Eq.~(\ref{eq:p1p1_ff}),
 large $\zeta$ implies that $\svxr~(\sim \lee g_X \cos\zeta \rii^4)$ 
becomes smaller. So $g_X$ has to increase to
achieve $\svxr \sim 10^{-19}$ cm$^3$sec$^{-1}$ for the X-ray line, 
as seen from the middle panel. In addition, larger $g_X$ implies
larger $\lan \sig v \ran_{\ga}$, which can be reduced by larger 
$\de$ (and a larger deviation away from the resonance region) as in the 
right-panel.\footnote{Again, $\svxr$ is insensitive to the $\de$ change in the saturated area as shown in Fig.~\ref{fig:sinx_sv}.}
Note that for the Majorana case, we do not spot this behavior for $g_X$ and 
$\de$ since $g_X$ is constrained to be less than $4\pi$.

To summarize, we present in Table~\ref{table:BMpts}
the chi-squares~($\chi^2$) and $p$-values for the best-fit points, and
also the $3\sig$ confidence regions for both the Majorana and Dirac DM
cases.  We would like to emphasize again that first, both cases can
explain the $\ga$-ray and X-ray data but only the Majorana DM can
reproduce the correct DM density. Second, the best-fit point shifts
toward the lower $m_{\chi_1}$ region when including the DM relic density
into the fits. Finally, $g_X$ is close to $4\pi$ in the Majorana case
because of $P$-wave velocity suppression as opposed to the Dirac case
where $g_X$ is of $\mathcal{O}(0.6)$.

\begin{table} 
\scriptsize
\centering
\begin{tabular}{| l | l |  l || l | }
\hline
\multicolumn{1}{|c|}{DM type}        &\multicolumn{2}{|c||}{Majorana}        & \multicolumn{1}{c|}{Dirac} \\
\hline
constraints & GC+Perseus &  GC+Perseus+$\Omega_{DM} h^2$ & GC+Perseus \\
\hline
\hline
DM mass (GeV)&  39.01 & 32.32 &28.6   \\
$\sin\chi$ &  $2.2\times 10^{-7}$ & $2.9\times 10^{-7}$ &$8.0\times 10^{-9}$  \\
$g_X$ &  12.56 & 12.43 &0.61   \\
$\delta$ &  $1.6\times 10^{-10}$ & $1.46\times 10^{-10}$ &$5.53\times 10^{-8}$  \\
$\chi^2$(GC)/$p$-value & 17.29/63.4\%  & 16.32/70\% & 15.53/74.5\%   \\
$\chi^2$(Perseus)/$p$-value & 3.36/33.9\% & 5.75/12.4\% & 0.023/99.99\%   \\
$\chi^2(\Omega_{DM} h^2)$/$p$-value & 20.40/0.01\% & 0.06/99.62\% &   100.0/0\%\\
\hline
\multicolumn{4}{|c|}{$99.73\%$ confidence region} \\
\hline
$m_{\chi_1}/m_{\psi_1}$ ($\gev$)&  [25.36, 57.90] & [24.38, 39.55] &[16.80, 56.11]   \\
$\sin\chi$ &  [$4.1\times 10^{-9}$, $1.4\times 10^{-6}$] & [$2.07\times 10^{-7}$, $3.86\times 10^{-7}$] &
[$1.45\times 10^{-10}$, $1.39\times 10^{-8}$] \\
$g_X$ &  [9.48, $4\pi$] & [9.71, $4\pi$] &[0.30, 1.48]   \\
$\delta$ &  [$10^{-10}$, $1.29\times 10^{-7}$] & [$10^{-10}$, $3.59\times 10^{-9}$] & 
[$8.3\times10^{-11}$, $1.09\times 10^{-6}$] \\
\hline
\end{tabular}
\caption{\label{table:BMpts} 
Best fit points of the Majorana and Dirac DM cases with
two sets of constraints:
GC+Perseus and GC+Perseus+$\Omega_{DM} h^2$. 
The best-fit point for the Dirac DM case are the same
for two sets of constraints.
The $p$-values are computed based on the degrees of freedom, 
3 for Perseus and $\Omega_{DM} h^2$ and 20 for GC.  
}
\end{table}

\section{Conclusions}\label{sec:conclusion}

In this work, we have attempted to explain simultaneously 
the GC $\ga$-ray excess and the 3.5 keV X-ray line,
as well as fulfilling the DM relic abundance
in the context of non-abelian DM models,
and we have success in the case of Majorana DM with a Dirac excited state.
We employed a ``dark'' $SU(2)_X$
gauge group with a $SU(2)_X$ doublet consisting of the DM particle and the
excited state with a mass-splitting of 3.5 keV. 
The $SU(2)_X$ sector talks to
the SM gauge groups via kinetic mixing, characterized by $\sc$.
We have studied two cases: Majorana DM~($\chi_1$) and Dirac DM~($\psi_1$),
and both with the Dirac excited state $\psi_2$. The X-ray line results
from $\chi_1 \chi_1~(\bar{\psi}_1 \psi_1) \to \bar{\psi}_2 \psi_2$,
followed by the decay of $\psi_2$ back to the DM and a photon. 
On the other hand, DM annihilations into SM fermions,
which then emit photons,
can explain the GC $\ga$-ray excess but this process is suppressed by
the aforementioned $SU(2)_X$-SM kinetic mixing.  In order to account
for the $\ga$- and X-ray data, one would need $\svxr \sim 10^{-19}$ and
$\lan \sig v \ran_{\ga}\sim 10^{-26}$ cm$^3$sec$^{-1}$.  
We employ the
resonance enhancement to fulfill the large $\svxr$. Additionally, the
large hierarchy between two cross-sections $\svxr / \lan \sig v \ran_{\ga}~(\sim 10^{7})$ can be realized
if the kinetic mixing is very small~($\sc \lesssim 10^{-7}$).

For the Majorana DM case, both the $\ga$-ray and $X$-ray excess can be
accommodated really well.  However, all $\chi_1$-involved processes
are $P$-wave suppressed due to the Majorana nature.  
As a result, the $SU(2)_X$ gauge coupling $g_X$
is driven close to the perturbativity limit $4\pi$ to counterbalance
the velocity suppression.  Regarding the DM relic density, the
relatively large DM velocity~($\sim\frac{1}{3} \, c$) at freeze-out compared to the
current one~($\sim 10^{-3}\, c$) implies a large deviation from the
resonance region.  Therefore, the cross-section of $\chi_1\chi_1 \to
\bar{f}{f}$ at freeze-out is much smaller than the current value
$10^{-26}$ cm$^3$sec$^{-1}$ demanded to explain the GC $\ga$-ray excess,
 as well as much smaller than 
$3\times10^{-26}$ cm$^3$sec$^{-1}$, the
size of the cross section to achieve the correct relic density. 
The solution comes from the $S$-wave dominated process
$\bar{\psi}_2\psi_2 \to \bar{f} f$, which should be included as an
coannihilation process at the freeze-out in light of the tiny mass
splitting between $\chi_1$ and $\psi_2$.  The coannihilation can reach
a level of $3\times 10^{-26}$ cm$^3$sec$^{-1}$ for certain
$m_{\psi_2}$ to reproduce the correct relic density.  The allowed
$m_{\chi_1}$~(also $m_{\psi_2}$) ranges from 25 to 40 GeV.  This coannihilation cross
section is much larger than the cross section of the $P$-wave process
$\chi_1\chi_1 \to \bar{f}{f}$.

In the Dirac DM case, the model can explain both $\ga$- and X-ray data 
for $16 \lesssim m_{\psi_1} \lesssim  56$ GeV, with  
much smaller $g_X~(\sim 0.6)$ compared to the Majorana DM case,
since all processes involved are $S$-wave dominated without the
velocity suppression.
Nevertheless, it cannot yield the proper DM relic density because 
both $\psi_1$ and $\psi_2$ annihilations into SM fermions are of the 
same order at freeze-out and both annihilations are away from the 
resonance region due to the large DM velocity. 
As a consequence, they are much smaller than $10^{-26}$
cm$^3$sec$^{-1}$, the current value for $\bar{\psi}_1 \psi_1 \to
\bar{f} f$, associated with the GC $\ga$-ray. An additional mechanism
has to be introduced to increase the annihilation cross-section and
lower the relic density. In addition, the direct search bounds hardly constrain our model since
the $SU(2)_X$-SM mixing $\sc$ is very small such that the DM-nucleon cross-section is negligible.

It is worthwhile to mention that recent studies~\cite{Bergstrom:2013jra,Ibarra:2013zia,Kong:2014haa},
based on the AMS-02 electron and 
positron data~\cite{Aguilar:2013qda,Accardo:2014lma},
infer stringent bounds on low-mass DM annihilation into leptons and
also quarks. In these works, the background with broken power-laws in energy fits the data very well, 
leading to strong constraints on the DM component. It might be arguable that the broken power-law background
is driven by the AMS02 data and somehow different from the conventional background.
Hence, we take a more conservative point of view that as long as predicted DM signals do not exceed the AMS-02
data. In this sense, such approach is not able to constrain our model.

Finally, we would like to comment that we have taken a
phenomenological approach, without justifying the smallness of the
mass splitting between the DM and the excited state, as well as
and the tiny kinetic mixing. Both are basically
determined based on the $\gamma$- and X-ray data.
Besides, as shown in Table~\ref{table:BMpts} for the $99.73\%$ confidence region,
the resonance condition of $m_{X^3}\simeq 2 m_{\chi_1}$
has to be precisely satisfied up to one part in $10^{9}$ or $10^{6}$ for the Majorana
or Dirac case respectively, where $\delta$ is driven to $10^{-9}$ for the strong enhancement on
the (co-)annihilation cross-section of the excited state to achieve the correct density in the Majorana case.
This fine-tuning results from the hierarchical annihilations required to
explain the X-ray line and GC $\gamma$-ray excess and it could arise from
an underlying flavor symmetry, aligned with the gauge symmetry such that
the resonance is not perturbed by large gauge couplings.   
The concrete model building is, however, beyond the scope of this work.

 \subsection*{Acknowledgments}

We would like to thank Andrew Frey for useful discussions.
W.~C. Huang is grateful for the hospitality of CERN theory group and
AHEP group in IFIC, where part of this work was performed. This work
is partially supported by the London Centre for Terauniverse Studies
(LCTS), using funding from the European Research Council via the
Advanced Investigator Grant 267352, and in part by
the MoST of Taiwan under Grant No. NSC 102-2112-M-007-015-MY3.
Y. Tsai was supported by World Premier International 
Research Center Initiative (WPI), MEXT, Japan.

\appendix

\section{Resonance enhancement and cancellation}
\label{reso_can}

We here show that if both the DM and the excited state are Majorana
particles and degenerate in mass, one cannot obtain
the resonance enhancement but the ``resonance cancellation'' instead. So it
is not capable of achieving the required large cross-section for the
X-ray line.  For demonstration, we use the following two types of
interactions, the vector current and axial-vector current, for
purely Dirac and Majorana DM:
\begin{align}
\mathcal{L}_M &\supset \sum^2_{i=1} g_X \chi_i^\dag \bar{\sig}^\mu \chi_i X_\mu, \nn \\
\mathcal{L}_D &\supset \sum^2_{i=1} g_X \bar{\psi}_i \ga^\mu \psi_i X_\mu,
\end{align} 
where $i=1~(2)$ corresponds to  DM~(excited state) 
with $\chi~(\psi)$ referring to a Majorana~(Dirac) particle.

\begin{figure}
   \centering
   \includegraphics[scale=0.5]{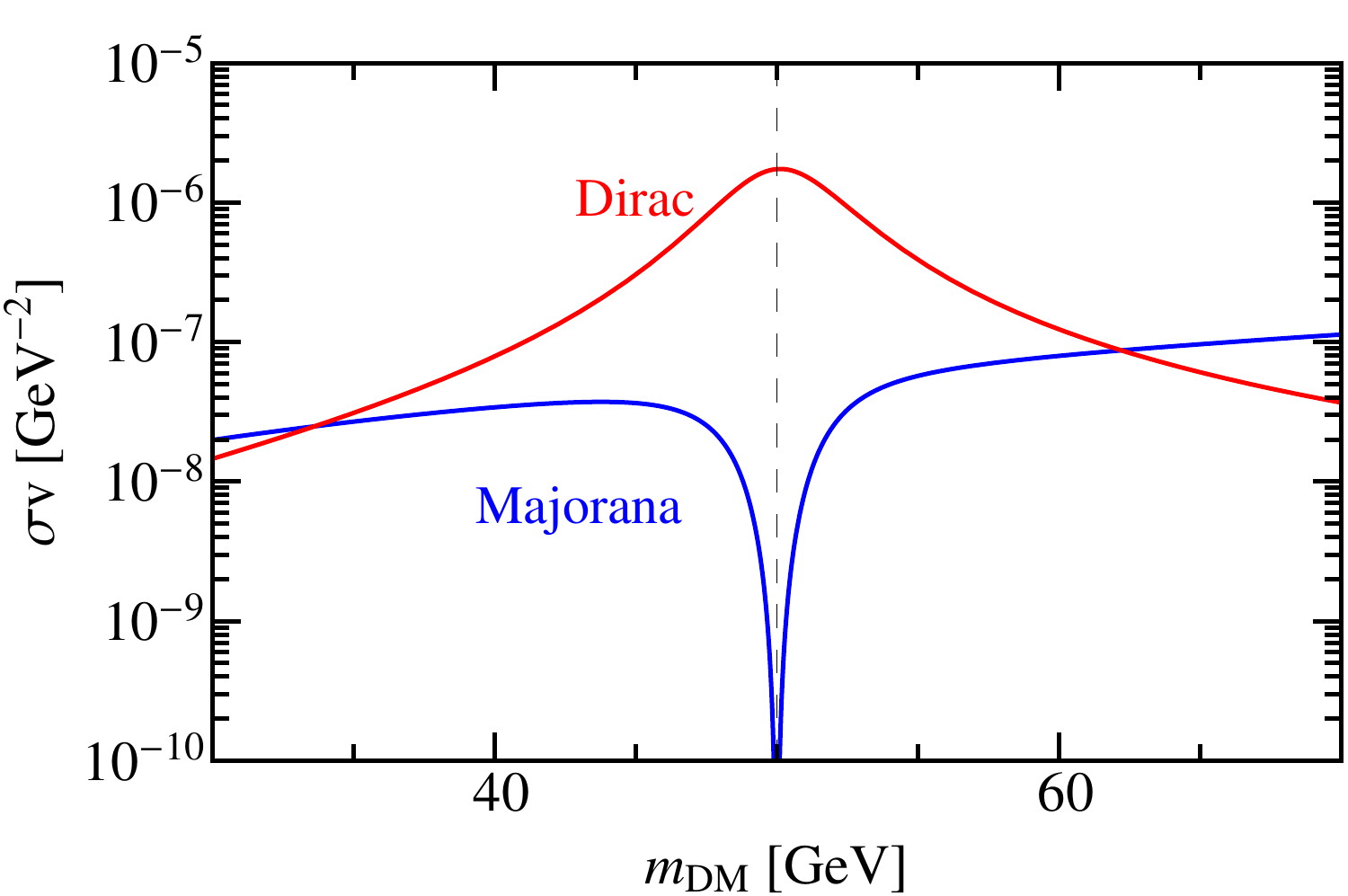}
 \caption{\emph{ The cross-section for the purely Majorana case~(blue) and Dirac case~(red). We assume a 3.5 keV mass splitting between DM and the excited state, $\Ga_X=\frac{m_X}{10}=10$ GeV and $v \sim 3\times 10^{-3} \, c$. }}
 \label{fig:D_vs_M_reso}
\end{figure}

In the vicinity of the resonance~($2m_{\chi(\psi)_1}\sim m_{X}$) and
the low DM velocity $v \ll c$, the resulting cross-sections for
$\chi_1\chi_1 \to \chi_2\chi_2$ and $\bar{\psi}_1\psi_1 \to
\bar{\psi}_2\psi_2$ through the current interactions read
\begin{align}
\lee \sig v \rii_{\chi_1\chi_1 \to \chi_2\chi_2} & \sim  \frac{ \lee m^8_{\chi_1} - m^6_{\chi_1} m^2_{\chi_2} \rii  v^2 +
12 m^4_{\chi_1} m^2_{\chi_1} \lee m_{X} - 2 m_{\chi_1} \rii^2 }
{ m^4_{X}  \lee \lee s -m^2_{X} \rii^2 + \Ga^2_{X} m^2_{X} \rii} , \nn \\ 
 \lee \sig v \rii_{\bar{\psi}_1\psi_1 \to \bar{\psi}_2\psi_2} & \sim  \frac{ 2 m^4_{\psi_1} +  m^2_{\psi_1} m^2_{\psi_2} }
{   \lee \lee s -m^2_{X} \rii^2 + \Ga^2_{X} m^2_{X} \rii},
\end{align} 
where we have suppressed coupling constants and coefficients from the
phase space integral and kinematics.  For the Majorana case, the first
term in the numerator is double suppressed because of $v \ll 1$ and
$m_{\chi_1} \simeq m_{\psi_2}$ and the second term becomes small in
the vicinity of the resonance. 
In contrast, in the Dirac case the
numerator is unsuppressed, which is characteristic of $S$-wave. In
Fig.~\ref{fig:D_vs_M_reso}, we show the comparison between the two cases,
in which we assume a 3.5 keV mass splitting between the DM and the excited
state, $\Ga_X=\frac{m_X}{10}=10$ GeV, and $v \sim 3\times 10^{-3} \,c$. 
It is clear that the pure Majorana case has the resonance
``cancellation'' instead of enhancement while the $S$-wave Dirac
case features the resonance behavior as expected.

\bibliography{GC_Xray}
\bibliographystyle{h-physrev}
\end{document}